\documentclass[a4paper,12pt]{article}
\usepackage{amsfonts,amssymb,amsmath}
\usepackage{bm,euscript,mathrsfs}
\usepackage[utf8]{inputenc}
\usepackage[T1]{fontenc}
\usepackage{color,soul}
\usepackage{textcomp}
\usepackage[tight,sf]{subfigure}
\usepackage{hyperref}
\usepackage{graphicx}
\usepackage{color}
\usepackage{psfrag}

\title{\Huge{Confinement of semiflexible polymers}}
\date{}
\author{\Large
Jemal Guven\footnote{\href{mailto:jemal@nucleares.unam.mx}{
jemal@nucleares.unam.mx}} \; and Pablo
V\'azquez-Montejo\,\footnote{\href{mailto:vazqmont@nucleares.unam.mx}{
vazqmont@nucleares.unam.mx}}}
\begin{document}
 \maketitle
\begin{center}
{\it Instituto de Ciencias Nucleares,
 Universidad Nacional Aut\'onoma de M\'exico\\
 Apdo. Postal 70-543, 04510 M\'exico, DF, MEXICO\\[1em]
}
\end{center}

\begin{abstract}
A variational framework is developed to examine the equilibrium
states of a semiflexible polymer that is constrained to lie on a
fixed surface. As an application the confinement of a closed polymer
loop of fixed length $2\pi R$ within a spherical cavity of smaller
radius, $R_0$, is considered. It is shown that an infinite number of
distinct periodic completely attached equilibrium states exist,
labeled by two integers: $n=2,3,4,\dots$ and  $p=1,2,3,\cdots$, the
number of periods of the polar and azimuthal angles respectively.
Small loops oscillate about a geodesic circle: $n=2$, $p=1$ is the stable ground state; states with higher
$n$ exhibit instabilities. If $R\ge 2R_0$ new states appear as
oscillations about a doubly covered geodesic circle; the state $n=3,
p=2$ replaces the two-fold as the ground state in a finite band of
values of $R$. With increasing $R$, loop states make a transition from  oscillatory
and orbital behavior on crossing the poles, returning to oscillation
upon collapse to a multiple cover of a geodesic circle (signaled,
respectively, by an increase in $p$ and an increase in $n$). The
force transmitted to the surface does not increase monotonically
with loop size, but does asymptotically. It behaves discontinuously
where $n$ changes. The contribution to energy from geodesic
curvature is bounded. In large loops, the energy becomes dominated
by a state independent contribution proportional to the loop size;
the energy gap between the ground state and excited states
disappears.
\end{abstract}

\section{Introduction}

Understanding how surfaces may
constrain the configuration of biopolymers on
mesoscopic scales is important in a number of processes in cell biology.
It  is particularly relevant in the packing of DNA within viral envelopes
%and cellular compartments
\cite{Spakowitz, Katzav}.
Modeling all of the relevant interactions is complicated: one needs to
accommodate the competition between polymer elasticity and entropy; in the case
of DNA, electric fields will be important. A nice short review in this  context
is provided in \cite{Kardar}. Recently, simulations treating various features of
the confinement process have  been performed \cite{Frey,Stoop2011}; the former
focuses on entropy, the latter on the dramatic effects of friction and  the
finite transverse dimensions of the confined object.

\vskip1pc \noindent
In this paper, we address an aspect of the problem  of a fundamental nature that
does not appear to have been addressed previously in any  detail: how does one
characterize the equilibrium states of the three-dimensional elastic
energy of a polymer confined within a fixed surface?  While this description of
confinement leaves out a lot of the physics that is relevant in biological
systems, it presents a well-defined problem exhibiting a striking level of
complexity that is worth studying in its own right.

\vskip1pc \noindent The semi-flexible polymer will be modeled as a
curve in three-dimensional space parameterized by arc-length $s$. The
bending energy associated with a given conformation is quadratic in
the Frenet-Serret curvature along the curve, $\kappa(s)$,
\begin{equation} \label{Hamk2}
H = \frac{1}{2}\int ds \,\kappa(s)^2\,.
\end{equation}
Curves of fixed length minimizing the unconstrained energy
(\ref{Hamk2}) were first studied in depth by Euler. A historical
review is provided in Ref. \cite{Levien}; a more contemporary
approach to the problem is presented in \cite{LangSing, IveySing}
and reviewed in Ref. \cite{SingerSantiago}. An alternative framework, which
lends itself to adaptation to the confinement problem, was developed in
Ref. \cite{Hamforcur}.

\vskip1pc \noindent
In the absence of constraints on a closed loop, there is not a lot to say: the
only stable equilibrium is a circle. Confinement within a surface, however, 
will generally oblige the loop to adopt a non-circular shape, increasing its
elastic energy. The contact with the surface itself may be complete or it may be
partial; and contrary to one's initial guess, the bound state will not
generally follow a surface geodesic; nor need it be unique.

\vskip1pc \noindent The wrapping of  a semi-flexible polymer around
a cylinder, a closely related problem relevant in the the winding of DNA around
histone octamers, was first examined some time ago by Nickerson and Manning
\cite{Mann, MannNick}.\footnote{See also \cite{MarkMann} and the work of Rudnick
and Zandi \cite{Rudnick}. A review is provided in Ref. \cite{ Schiessel}.
There has also been some nice work done more recently by Van der Heijden et al.
\cite{Dutch}. More directly relevant is the study of confinement of cylindrical
sheets within a circular cylinder by Bou\'e et al. \cite{Boudaoud}.} Their
strategy was to look at the independent degrees of freedom of the
surface-bound polymer.  While focusing directly on these degrees of
freedom makes sense, it does not exploit the symmetries of the
problem. For, even though the constraint breaks the Euclidean invariance
of the three-dimensional bending energy, how this occurs is not
arbitrary.  A variational framework is developed here that
involves the unconstrained degrees of freedom, imposing the
constraint  using a local Lagrange multiplier. This multiplier will
quantify the loss of Euclidean invariance in the constrained system.
Its value at any point along the loop will be identified as the
local normal force that is being transmitted to the surface.

\vskip1pc \noindent
The well-known integrability of the Euler-Lagrange equations for the
unconstrained curve is a  consequence of the Euclidean invariance of its energy.
The constrained counterpart generally will not be integrable. In various
interesting cases, however, the confining geometry will respect some
subgroup of the Euclidean group. In particular, the conservation of
torque associated with the residual rotational invariance of a
sphere permits the Euler-Lagrange equation to be cast as a quadrature in the
geodesic curvature which can be integrated directly. It also provides a direct
recipe for the reconstruction of the loop from its curvature data adapted to the
conserved torque vector as an axis of symmetry.

\vskip1pc \noindent
In particular, we apply this framework to examine the  confinement of a
closed polymer loop of fixed length $2\pi R$ within a sphere of radius $R_0\le
R$.

\begin{figure}
\begin{center}
  \subfigure[$ R = 1.1$]{\includegraphics[scale=0.12]{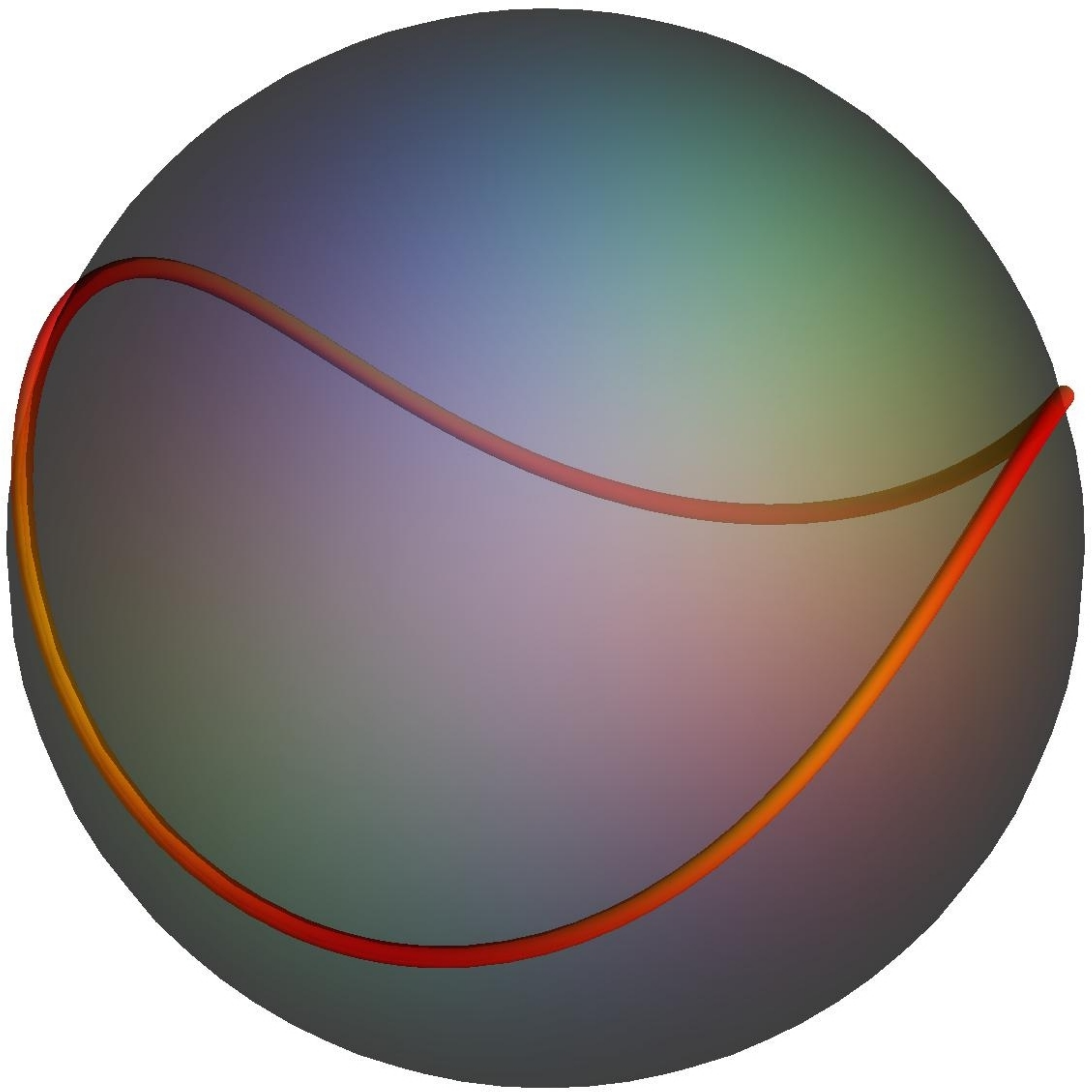}}
  \hfill
  \subfigure[$R = 1.5$]{\includegraphics[scale=0.12]{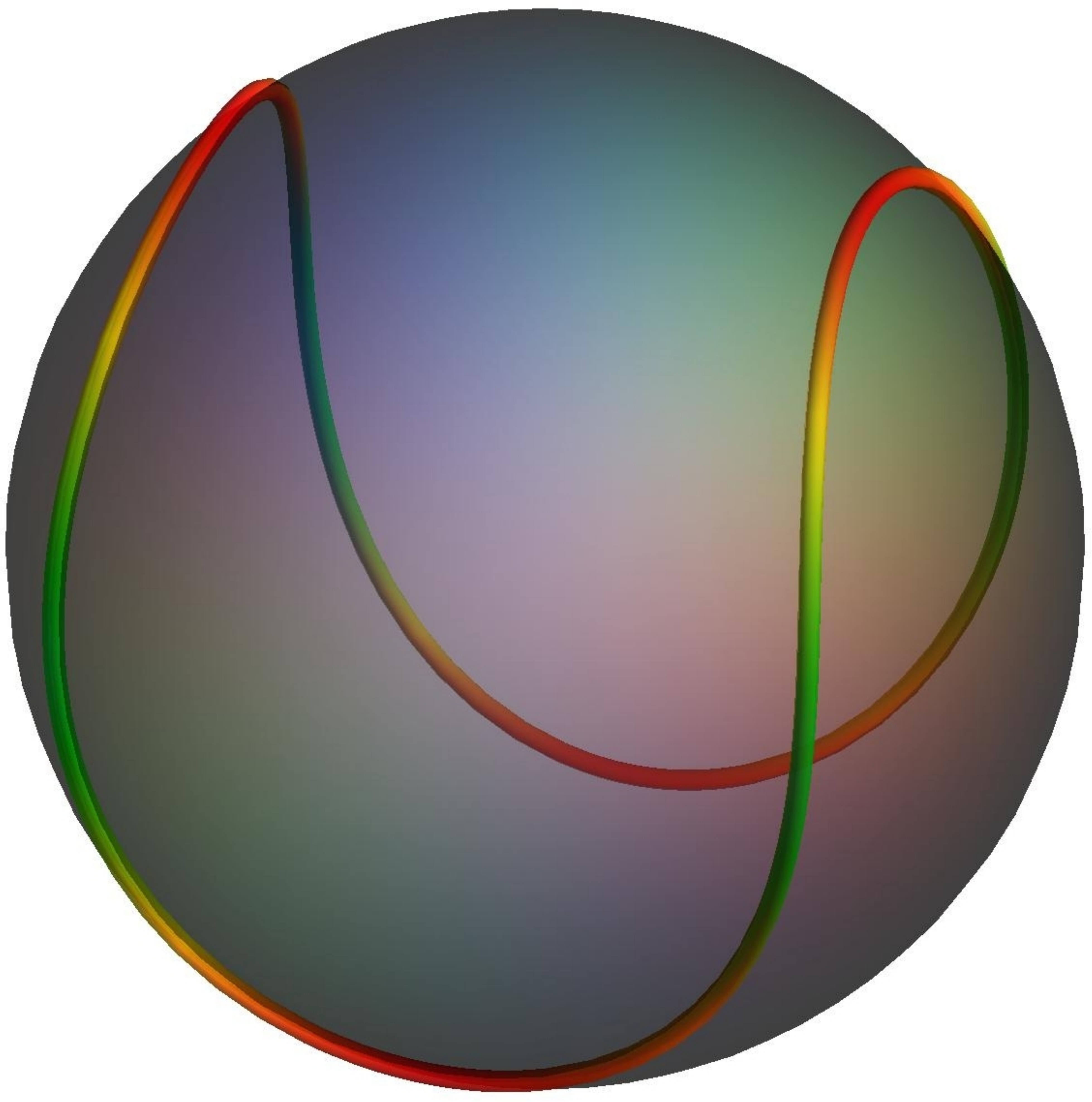}}
  \hfill
  \subfigure[$R = 2$]{\includegraphics[scale=0.12]{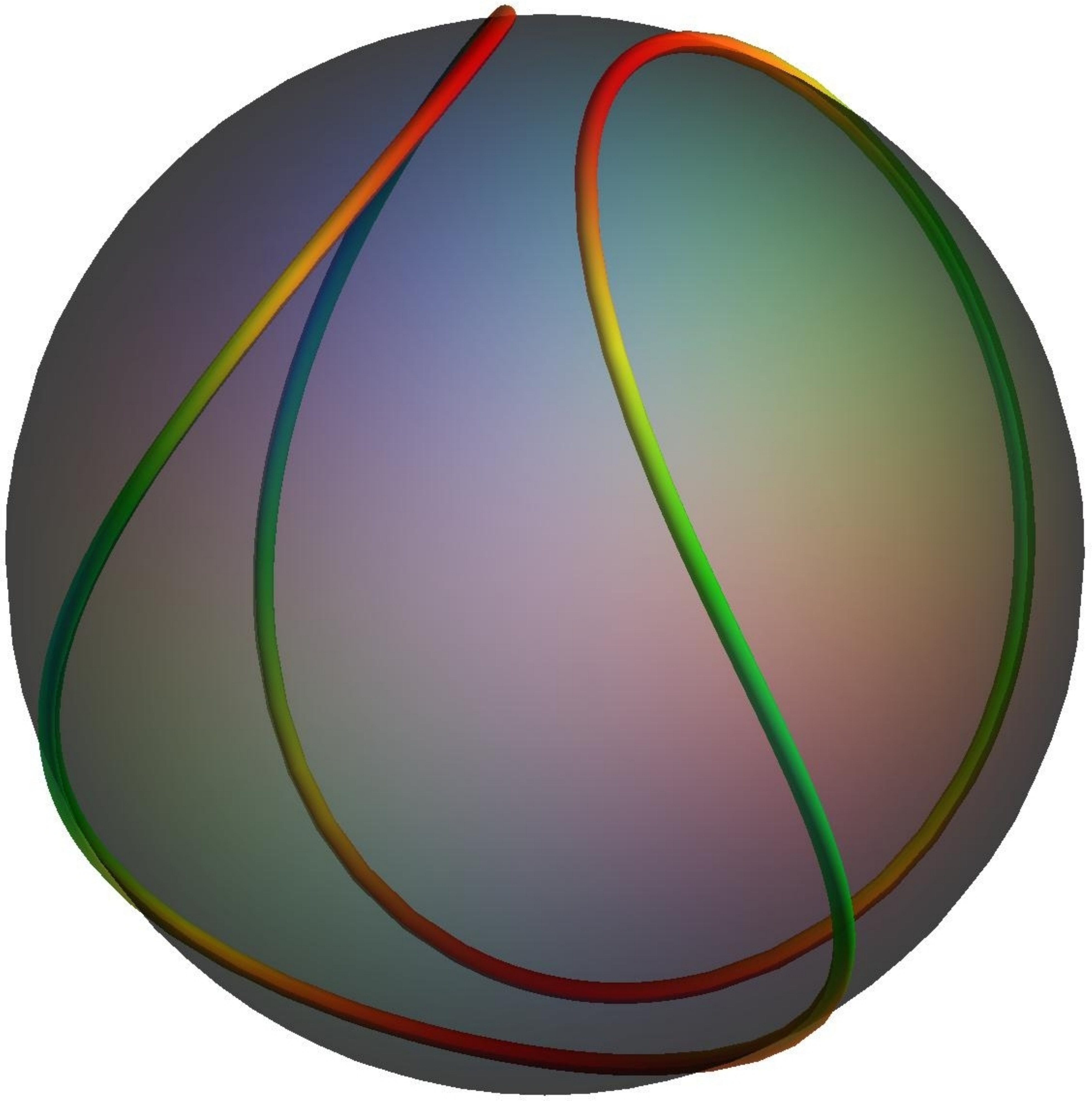}}\\
  \subfigure[$R = 2.127$]{\includegraphics[scale=0.12]{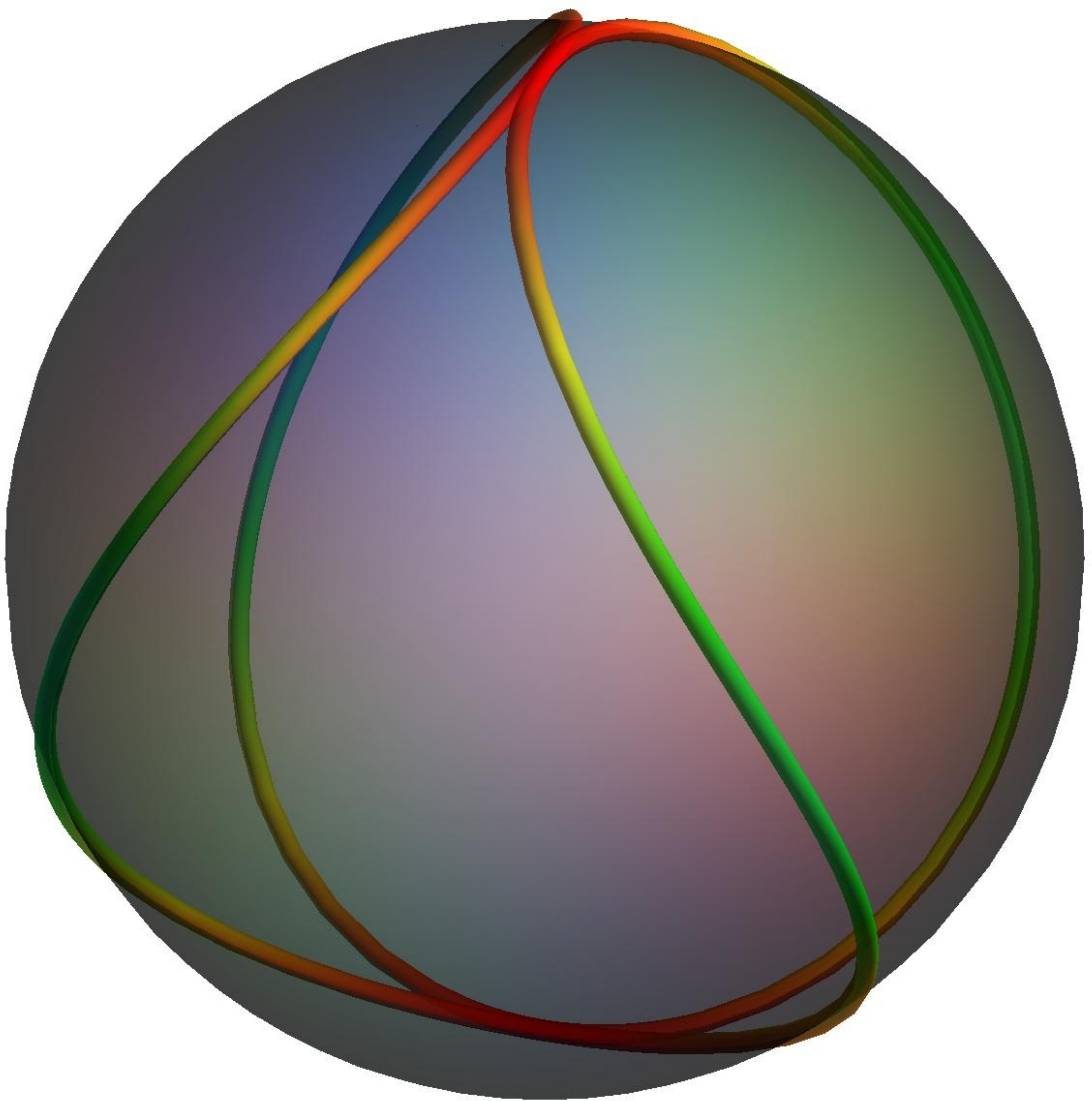}}
  \hfill
  \subfigure[$R = 2.5$]{\includegraphics[scale=0.12]{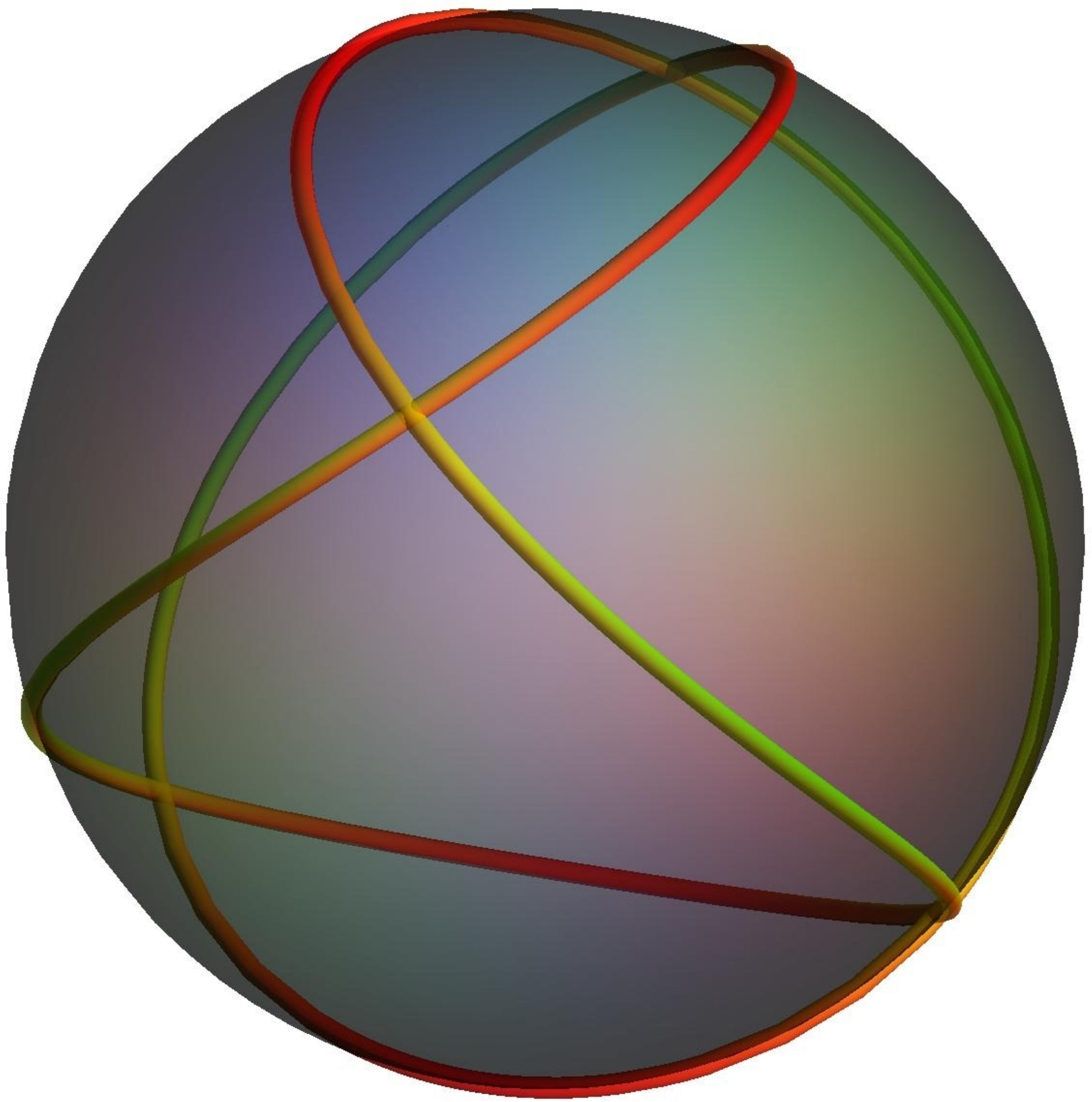}}
  \hfill
  \subfigure[$R = 2.99$]{\includegraphics[scale=0.12]{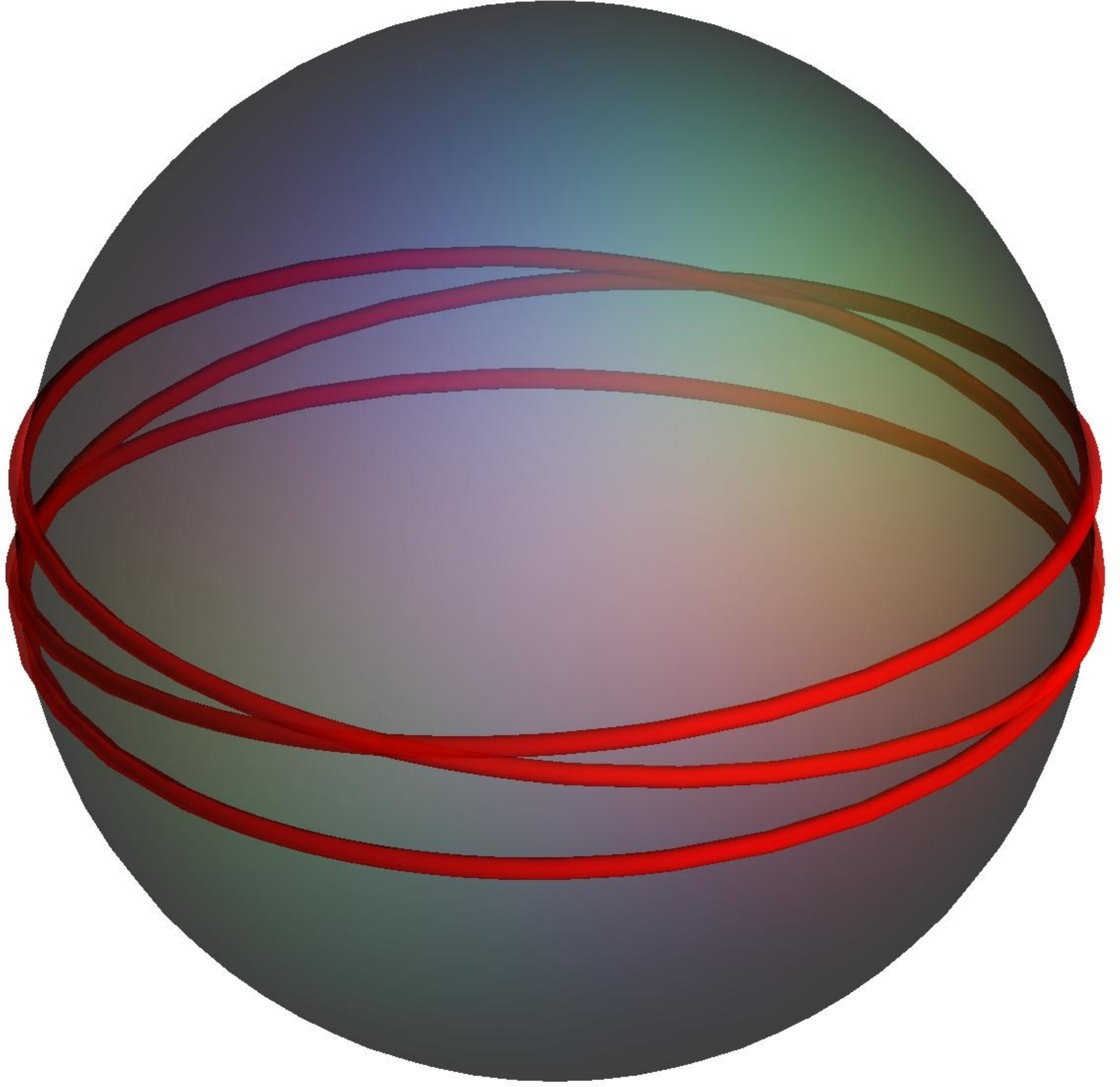}}\\
\includegraphics[scale=0.175]{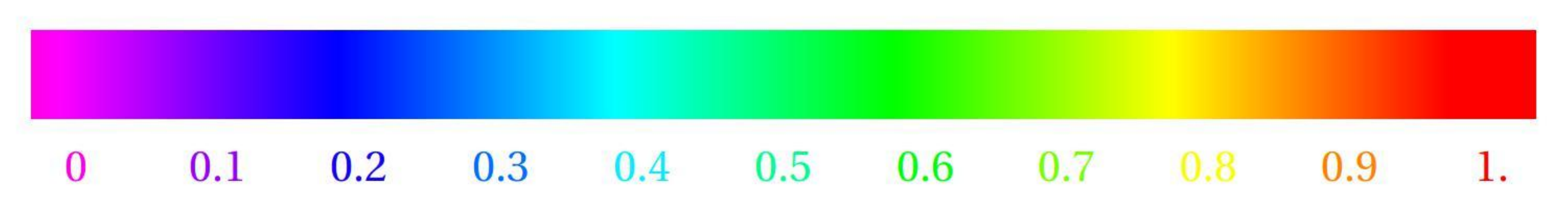}
\end{center}
\caption{\small (Color online) State with two-fold symmetry for values of $R$ in
the interval $[1,3]$ within a unit sphere: (a) and (b) display increasingly large
oscillations about the equator; (c) oscillations develop
overhangs; (d) first self-contacts made at the two poles;  (e)
self-intersecting triple orbit of sphere; (f) orbit collapses to
triple cover of the equator. The normalized local confining force
$\lambda$ is color coded in these figures.}\label{figure1}
\end{figure}

\vskip1pc\noindent In contrast to an open polymer which will wind
around a geodesic circle on the sphere when its length exceeds $\pi
R_0$, the closure of the loop is incompatible with a geodesic unless
$R$ is tuned to be commensurate with an integer multiple of $R_0$.
We show that there exists an infinite number of distinct
completely attached states, labeled by a pair of integers, $n$ and
$p$: $n=2,3,4\dots$, $p=1,2,3,\dots $, the number of periods of the
polar angle and azimuthal angles in one circuit of the loop
respectively. $n$ characterizes the dihedral symmetry with respect
to the axis of symmetry; $p$ characterizes the number of revolutions about this
axis. Small loops oscillate symmetrically about a geodesic circle
with $p=1$ and an $n$-fold symmetry, $n=2,3,\dots$. The two-fold is
the stable ground state. For any finite values of $R$, the higher
$n$-folds are unstable with respect to decay toward the two-fold.
States with $n=2$ are illustrated in Figs. \ref{figure1}(a) - \ref{figure1}(c) 
for various normalized values of $R$.

\vskip1pc \noindent
Beyond some critical size, the loop will exhibit self-intersections on the
sphere.  When $n=2$ this will occur when $R=2.127 R_0$ where the loop crosses
the poles as illustrated in Fig.  \ref{figure1}(d). We suppose that
self-intersections are consistent with the physics and do not cost energy.

\vskip1pc \noindent As $R$ is increased to $R=3R_0$, the loop will
collapse onto a geodesic circle which it will cover three times, (see
Figs. \ref{figure1}(e) and \ref{figure1}(f). The discontinuity in the number of
revolutions occurs at an intermediate value of  $R$ where the loop crosses the
poles. This is accompanied by a transition from oscillatory to orbital behavior.
As $R$ is increased above $3R_0$, the dihedral symmetry of the lowest energy
state descending from the $n=2$ ground state will jump to $n=4$ with the
reestablishment of oscillatory behavior.

\vskip1pc \noindent
The bending energy of this state as well as the total force that gets
transmitted to the surface will be determined as a function of $R$. The three-
dimensional bending energy decomposes on the surface into a sum of two terms:
one is associated with the geodesic curvature, intrinsic to the sphere;
the other is associated with the normal curvature inherited from the surface,
constant on a sphere. The latter energy thus  counts the number of times the
loop is wound within the sphere and it is proportional to $R$. The geodesic
energy periodically falls to zero whenever $R=pR_0$ and the loop is geodesic.
Local maxima, associated with the incommensurability of the loop length with 
geodesic behavior, are displayed between these values; we show that their
values decrease monotonically with loop size. For large values of $R$, the
normal contribution is always dominant. It does not depend on the state in
question. Thus, in this limit, the energy gap between the ground state and
excited states disappears. The local force transmitted by the completely
attached ground state loop is positive everywhere. Surprisingly, the total force
does not grow monotonically with loop length except asymptotically, where it
grows linearly with $R$ and coincides with the  naive expression, energy divided
by $R_0$.  The change of symmetry as $R$ passes through odd integral multiples
of $R_0$, manifests itself in a positive jump in the transmitted force analogous
to a Euler instability associated with the buckling into oscillations.

\vskip1pc \noindent When $R\ge 2R_0$, a new set of states appears as
oscillations about a doubly covered geodesic circle, $p=2$ with an
$n$-fold symmetry, $n=3,4,5,\dots$; among these new states, the lowest
energy is displayed in the three-fold. Also, because the geodesic
energy is small when the oscillations are small, these states will
have lower energy than the two-fold ground state in a finite band of
values of $R$ beginning at $2R_0$. There will be similar behavior in
an infinite set of bands of values of $R$ where the descendant
states become geodesic.  One thus needs to reassess the stability of
the two-fold ground state and its descendants. We provide a
heuristic argument for stability  by constructing a homotopy that
interpolates between the states with $n=2$, $p=1$ and $n=3$, $p=2$.
By examining the energy along this homotopy, we show that a steep
energy barrier separates the two equilibrium states. The $n=2$,
$p=1$ state and its descendants thus appear to be stable
classically. For any $R\ge 2R_0$, there are two stable states  that
alternate as the ground state as the length is increased.

\vskip1pc \noindent
The paper is organized as follows: In section \ref {confinedelastica}, we
describe the framework. In particular, the breaking of translational invariance
will be quantified by the non-conservation of a vector along the loop. It will
be shown that the Euler-Lagrange equation  for the  curve can be cast as the
vanishing of a linear combination of the unconstrained Euler-Lagrange
derivatives. The constraining force is identified as some other linear
combination of these derivatives. In Sect. \ref{Spherical} the confinement of
a closed loop by spheres will be considered. In Secs. \ref{weak} and
\ref{Strong} we analyze loops in the harmonic and non-linear regimes
respectively. The equilibrium states of the loop will be identified and the
forces they transmit to the surface determined. An assessment of the
stability of these states will be provided. We conclude with a discussion and a
few suggestions for future work in Sec. \ref{Discussion}. A number of useful
definitions, identities and derivations are collected in a set of appendixes.

\section{Curves constrained to surfaces} \label{confinedelastica}

Consider a space curve $\Gamma: s \rightarrow \mathbf{Y}(s)$
parametrized by arclength constrained to lie on a surface
$\Sigma$.  This surface is described in parametric form by the
mapping $\Sigma: (u^1,u^2) \rightarrow \mathbf{X}(u^1,u^2)$. The
confined curve can then also be described as a surface curve
$\Gamma_\Sigma: s\rightarrow (U^1(s),U^2(s))$. In order to enforce
the condition that $\Gamma$ lie on $\Sigma$, one adds to the
energy $H[\mathbf{Y}]$ given by Eq. (\ref{Hamk2}) a term enforcing
this constraint:
\begin{equation} \label{confconst}
H_c [\mathbf{Y}, U^a] = H[\mathbf{Y}] + \int \, ds \,\, \bm{\lambda}(s) \cdot
\left[ \mathbf{Y}(s) -
\mathbf{X} (U^a(s)) \right]\,,
\end{equation}
where $\bm{\lambda}$ is a vector-valued Lagrange multiplier
defined along the curve. This constraint will break the manifest
translational invariance of the energy $H$.

\vskip1pc \noindent
The variation of $H_c$ with respect to the embedding functions $\mathbf{Y}$ can
be cast in the form
\begin{equation}
\delta_{\bf Y} H_c = \int ds \left(\mathbf{F}' + {\bm \lambda}\right) \cdot
\delta \mathbf{Y}\,,
\end{equation}
where prime represents derivation with respect to arclength and the tension in
the loop ${\bf F}$ is  given by
\begin{equation} \label{EEFSforce}
\mathbf{F} = \left(\frac{1}{2} \kappa^2 - c \right) \mathbf{T} + \kappa'
\mathbf{N} + \kappa \tau
\mathbf{B}\,.
\end{equation}
Here $\{{\bf T},{\bf N},{\bf B}\}$ is the standard Frenet-Serret
frame carried by the curve and $\tau$ is its torsion. The constant
$c$ is associated with the constraint of fixed length. A
derivation of Eq. (\ref{EEFSforce}) is provided in \ref{Stresstender}.

\vskip1pc \noindent In equilibrium, one finds that  $\mathbf{F}' =
- \bm{\lambda}$. Thus, in the presence of the constraint, the
tension in the loop is not conserved; the multiplier is identified
as the external force associated with the constraint
\cite{LandauLifshitz}.

\vskip1pc \noindent
The corresponding variation of $H_c$ with respect to $U^a(s)$ is given by
\begin{equation}
\delta_U H_c = - \int ds \,\bm{\lambda} \cdot \mathbf{e}_a\, \delta U^a\,,
\end{equation}
where ${\bf e}_a$, $a=1,2$ are the two tangent vectors to the
surface adapted to the parametrization by $u^a$. In equilibrium,
$\bm{\lambda}\cdot {\bf e}_a =0$ ; in equilibrium, the force on
the curve associated with the constraint always acts orthogonally
to the surface.  Let us write  $\bm{\lambda} = \lambda
\,\mathbf{n}$, where $\mathbf{n}$ is the unit vector normal to
$\Sigma$. The combined result is that
\begin{equation} \label{EL}
\mathbf{F}' = - \lambda \mathbf{n}\,.
\end{equation}

\vskip1pc \noindent
An integrability condition for closed curves follows from Eq. (\ref{EL}):
\begin{equation}
\oint ds \, \lambda\, {\bf n}=0\,.
\end{equation}
This identity holds whether or not contact is complete.

\vskip1pc\noindent
Using Eq. (\ref{EEFSforce}) along with the Frenet-Serret equations, a
straightforward calculation decomposes ${\bf F}'$
along the normals
\begin{equation}\label{eq:FprimeEL}
{\bf F}' =
\varepsilon_{\bf N} \mathbf{N} + \varepsilon_{\bf B} \mathbf{B}\,,\end{equation}
where the Euler-Lagrange
derivatives of the bending energy $\varepsilon_{\bf N}$ and $\varepsilon_{\bf
B}$ are given by
\begin{subequations}
\begin{eqnarray}
\varepsilon_\mathbf{N}&=&\kappa^{''} +
\kappa\, \left( \frac{\kappa^2}{2} - \tau ^2 - c \right)
\,,\label{ELELn}\\
\varepsilon_\mathbf{B}&=& \frac {2}{\kappa} \,
\left(\kappa^2 \tau \right)' \,.\label{ELELb}
\end{eqnarray}
\end{subequations}
The tangential Euler-Lagrange derivative $\varepsilon_{\bf T}=
\mathbf{F}' \cdot \mathbf{T}$ vanishes identically, a consequence
of the fact that the only relevant degrees of freedom are
geometrical, whether the curve is constrained or not.

\vskip1pc \noindent
The surface-bound curve also carries a Darboux frame,
$\{{\bf T},{\bf n}, {\bf l}= {\bf T}\times {\bf n}\}$.
Relevant properties of this frame are summarized in \ref{AppDarbouxframe}.  The
Frenet normals are related to their Darboux counterparts by a rotation about the
tangent vector:
\begin{equation}
{\bf N}= \cos\omega\, {\bf n} + \sin\omega  \,{\bf l}\,;\quad
{\bf B}= - \sin\omega \, {\bf n} + \cos\omega\, {\bf l}\,.
\end{equation}
The relationship ${\bf T}'=\kappa {\bf N}$ permits one to express the
angle of rotation in terms of a curvature ratio: $\sin\omega = \kappa_g/\kappa$
or $\cos\omega = \kappa_n/\kappa$, where $\kappa_g$  and $\kappa_n$ are,
respectively, the geodesic and normal curvatures along $\Gamma_\Sigma$,
\begin{equation}
\kappa_g = {\bf T}'\cdot {\bf l} \,,\quad \kappa_n= {\bf T}' \cdot {\bf n}\,.
\end{equation}
Using these expressions it is possible to decompose ${\bf F}'$ given by Eq.
(\ref{eq:FprimeEL}) in a form adapted to the surface,
\begin{equation}
{\bf F}' = \left( \cos\omega \,\varepsilon_{\bf N} - \sin\omega
\,\varepsilon_{\bf B}
\right)\mathbf{n} + \left(\sin\omega \,\varepsilon_{\bf N} + \cos\omega\,
\varepsilon_{\bf B}\right)
\mathbf{l} \,.
\end{equation}
The projection of  Eq. (\ref{EL}) onto $\mathbf{l}$  provides the Euler-Lagrange
equation in the remarkably simple form,
\begin{equation} \label{eulelsurfeneb}
 \varepsilon_\mathbf{l} = \sin\omega \,\varepsilon_\mathbf{N} + \cos\omega
\varepsilon_\mathbf{B} =
0\,.
\end{equation}
The multiplier $\lambda$ does not appear. The corresponding projection onto
${\bf n}$ determines $\lambda$,
\begin{equation} \label{magnlambda}
\lambda = - \cos \omega  \, \varepsilon_\mathbf{N} + \sin\omega  \,
\varepsilon_\mathbf{B}  \,.
\end{equation}
In this approach, one sees explicitly how both the Euler-Lagrange equation
(\ref{eulelsurfeneb}) and the confining force (\ref{magnlambda}) are constructed
out of the two unconstrained Euler-Lagrange derivatives. The normal force is
completely determined  when the local geometry is known.

\vskip 1pc \noindent The Euler-Lagrange equations for an
unconstrained elastic curve, given by $\varepsilon_{\bf N} = 0$
and $\varepsilon_{\bf B} = 0$, are replaced by the single
equation, $\varepsilon_{\bf l} =0$. The apparent  discrepancy in
the number of equations reflects the fact that a space curve
possesses two  independent modes of deformation whereas a surface
curve has only one. In general, the integrability of the former
pair of equations is surrendered when the constraint is present.

\subsection{Euler-Lagrange equation in terms of surface curvatures}
\label{sect:confeulerelastica}

Using the identities  (\ref{kaptau}) and (\ref{dkappagn}) the
Euler-Lagrange equation (\ref{eulelsurfeneb}) can be expressed
completely in  terms of surface curvatures,
\begin{equation} \label{ELEulerelastonasurf}
\varepsilon_\mathbf{l} =\kappa_g''+\kappa_g \left(\frac{\kappa_g^2 +
\kappa_n^2}{2} -
\tau_g^2-c\right) -\frac{\left(\kappa^2_n \tau_g\right)'}{\kappa_n} = 0\,.
\end{equation}
This agrees with the equation derived in Ref. \cite{MannNick}
using a very different approach. Note that it involves the
curvatures as well as the geodesic torsion and, in general, the
curve will not follow a geodesic with $\kappa_g =0$. For a
geodesic to minimize bending energy, one requires one of the
following to occur: $\kappa_n^2 \tau_g$ is constant; the curve
coincides with an asymptotic line with $\kappa_n=0$ or a principal
curve with $\tau_g=0$. Such conditions typically do not occur
unless they do so trivially.

\vskip 1pc \noindent
The magnitude of the force $\lambda$ transmitted to the surface, given by Eq.
(\ref{magnlambda}), assumes the form
\begin{equation} \label{CEELmagnlambda}
-\lambda = \kappa_n'' + \kappa_n \left( \frac{\kappa_g^2 + \kappa_n^2}{2} -
\tau_g^2
-c\right) + \frac{\left(\kappa_g^2 \tau_g \right)'}{\kappa_g}\,.
\end{equation}
Its magnitude will vary along the contact region, even for
confinement by a sphere. This expression is missing in the
framework presented in \cite{MannNick}. A curious consequence of
the symmetric decomposition of the Frenet-Serret curvature in
terms of the geodesic and the normal curvatures is the fact that
$\lambda$  turns out to be identical to the Euler-Lagrange
derivative $\varepsilon_\mathbf{l} $ given by
(\ref{ELEulerelastonasurf}) under the interchange of  $\kappa_g$
with  $\kappa_n$ and a change of sign of the last term.

\subsection{Confinement and the loss of rotational invariance}
\label{sect:rotations}

\vskip1pc \noindent In general, under confinement,  one surrenders
not only translational invariance but also rotational invariance.
The torque about the origin per unit length of the curve, ${\bf
M}$, is  given by
\begin{equation} \label{eq:MFS}
 {\bf M}= \ {\bf Y}\times {\bf F} + {\bf S}\,,
\end{equation}
where ${\bf S} = -\kappa {\bf B} = \kappa_g \, {\bf n} - \kappa_n
\, {\bf l}$ . The first term on the right in Eq. (\ref{eq:MFS}) is
the torque due to the force ${\bf F}$; the second term is the
bending moment originating in second derivatives in the bending
energy. For a free  curve, ${\bf M}'=0$, which can be cast in the
manifestly translationally invariant form  $\mathbf{S}'+
\mathbf{T}\times \mathbf{F}=0$ \cite{LandauLifshitz}.
 In general, ${\bf M}$ is not conserved. One has instead
\begin{equation}
\mathbf{M}' = \varepsilon_\mathbf{l} \left(\mathbf{Y}\times \mathbf{l}
\right) - \lambda \left(\mathbf{Y} \times\mathbf{n}\right) \,.
\end{equation}
Thus, in a confined equilibrium with $\varepsilon_\mathbf{l}=0$, ${\bf M}$ will
not generally
be conserved. The source is  given by the moment of the force associated with
the constraint.

\vskip1pc \noindent
If the confining geometry is a sphere
centered on the origin, so that $ \mathbf{Y}$ is directed along the normal
vector,
${\bf M}$ will be conserved. If it is symmetric about some axis (say the $z$
axis),
then the conserved quantity is the corresponding projection of ${\bf M}$, that
is, $M_3 = {\bf M}\cdot \hat{\bf z}$.

\section{Spherical confinement} \label{Spherical}

Consider a closed curve of length $S = 2 \pi R$, confined within a
sphere of radius $R_0$. We normalize lengths in terms of
$R_0$. On a sphere, the extrinsic curvature tensor is proportional
to the metric $K_{ab} = g_{ab}$, so that the  normal curvature is
constant, $\kappa_n = - 1$, and the geodesic torsion vanishes,
$\tau_g = 0$.  Using Eq. (\ref{eq:CEEF}), the vector $\mathbf{F}$
given by Eq. \ref{EEFSforce}, then reduces to
\begin{equation} \label{CEEFSphere}
\mathbf{F} = \left(\frac{\kappa_g^2}{2} + \sigma \right) \mathbf{T} + \kappa_g'
\mathbf{l}\,\qquad
\text{where} \qquad \sigma = \frac{1}{2 } - c\,.
\end{equation}
${\bf F}$ is everywhere tangent to the surface. It is also completely determined
by the intrinsic geometry. However, it is not conserved.
\vskip1pc \noindent
The torque vector defined by  Eq. (\ref{eq:MFS}) is given by
\begin{equation} \label{Torquevctsphcnf}
 \mathbf{M} = \kappa_g' \mathbf{T} -\left(\frac{\kappa_g^2}{2} + \sigma -
1\right) \mathbf{l} + \kappa_g \mathbf{n}\,.
\end{equation}
The rotational invariance of bending energy confined to a sphere
implies that ${\bf M}$ is a constant vector. In particular, its
length is a constant. This provides a quadrature for the geodesic
curvature  ($ M^2 = {\bf M}\cdot {\bf M} $):
\begin{equation} \label{CEEsphereM2}
M^2 = (\kappa_g')^2 + \left(\frac{\kappa_g^2}{2} +\sigma - 1 \right)^2 +
\kappa_g^2\,.
\end{equation}
It is simple to check that the condition $(M^2)'=0$  in
Eq. (\ref{CEEsphereM2}) reproduces the Euler-Lagrange
Eq. (\ref{ELEulerelastonasurf}) on a sphere:
\begin{equation} \label{CEELsphere}
\varepsilon_{{\bf l}} = \kappa_g'' + \kappa_g \left(\frac{\kappa^2_g}{2} +
\sigma \right) = 0\,.
\end{equation}
This equation is also identical to the one describing an elastic
loop on a sphere with an energy density, $\kappa_g^2/2$. The
minimization of the constrained three-dimensional bending energy
coincides in this case with that of  two-dimensional spherical
bending energy. This is a consequence of the decomposition of the
Frenet-Serret curvature into normal and geodesic parts,
$\kappa^2=\kappa_g^2+\kappa^2_n$ as well as  the fact that
$\kappa_n$ is a constant.  Spheres are special in this respect.

\vskip1pc \noindent Various mathematical properties of elastic curves on spheres
were described by Langer and Singer in the 1980s \cite{LangSing}; see also
\cite{Crouch} for  a numerical treatment of the problem.  Recently they have
been reexamined in some detail in \cite {Arroyo}. The connection to the
description of conical defects in unstretchable flat sheets was developed in
\cite{JemalMartin} and \cite{JemalMartinBenAmar}. While the mathematical
literature provides a  useful point of departure, the absence of any reference
either to energy or to the transmitted forces
limits its  usefulness.

\vskip1pc \noindent
Note that in a geodesic, $\kappa_g=0$ everywhere. Equation (\ref{CEEsphereM2})
then implies that $M^2 = (\sigma-1)^2$.  Unless $R$ is an integer, however,
geodesics will be inconsistent with the boundary conditions associated with
closure. If $\kappa_g=0$ is accessible anywhere along the loop, there will be a
non-trivial bound on $M$ from below:
\begin{equation}
\label{eq:parabola}
(\sigma-1)^2 \le  M^2 \,.
\end{equation}
In equilibrium, all confined loops will satisfy this bound.

\vskip1pc \noindent
It is also useful to cast Eq. (\ref{CEEsphereM2})
in the alternative form
\begin{equation} \label{CEEspherefirstint}
(\kappa_g')^2 + V(\kappa_g) = E^2\,,
\end{equation}
where
\begin{equation} \label{eq:Vdef}
V(\kappa_g) = \left(\frac{\kappa_g^2}{2} +\sigma \right)^2\,,
\end{equation}
and $E^2 = M^2 + 2 \sigma -1$ is manifestly positive.\footnote {This
is a weaker bound than Eq. (\ref{eq:parabola}).}

\vskip1pc \noindent If $\kappa_g$ is identified as the position of
a particle and $s$ as time, then Eq. (\ref{CEEspherefirstint}) described
the motion of a particle of mass $m=2$ and  total ``energy'' $E^2$
in the symmetric quartic potential $V(\kappa_g)$.  If $\sigma\ge
0$, $V$ possesses a single minimum at $\kappa_g=0$; if $\sigma<0$
it possesses two symmetric wells centered at $\kappa_g = \pm
\sqrt{-2\sigma}$, separated by a local maximum at $\kappa_g=0$.

\vskip1pc \noindent
While the particle analogy is useful, it does
have its limitations. $E^2$ also is not the energy of the loop and
the potential depends on the constant of integration $\sigma$
which we are not free to tune but, like the `energy', is itself
determined by the boundary conditions.

\vskip1pc \noindent The qualitative behavior of the loop will
depend on the turning points of the potential, and thus on the
relative values of $\sigma^2$ and $E^2$:

\vskip1pc \noindent
(1) $E^2 >\sigma^2$ (equivalently Eq. (\ref{eq:parabola}) is satisfied).
In this parameter regime, there are only two turning points.
One has
\begin{equation}
\label{eq:turningpoints}
E^2 - V(\kappa_g) = \frac{1}{4} \left(\kappa^2_g + K^2\right) \left(k_1^2
 -
\kappa^2_g\right)\,,
\end{equation}
where $K^2 = 2 (E + \sigma) \geq 0$ and $k_1^2 = 2 (E - \sigma)
\geq 0$. Thus $\kappa_g'=0$ when $\kappa_g = \pm k_1$.
The geodesic curvature thus ranges in the symmetric interval $[-k_1,k_1]$. This
will be independent of the sign of $\sigma$. The loop will
oscillate symmetrically about the equator where $\kappa_g=0$ (see
Eq. (\ref {Eq:theta})) so that $\oint ds \kappa_g = 0$.

\vskip1pc \noindent
(2) $E^2< \sigma^2$. This regime is inaccessible physically if the loop is
closed.\footnote{ It can only arise if $\sigma<0$ so that the potential
possesses two wells and the trajectory in $\kappa_g$ is confined to oscillate in
one of them. One can write
\begin{equation}
E^2 - V(\kappa_g) = \frac{1}{4} \left(\kappa^2_+ - \kappa^2_g\right)
\left(\kappa^2_g -
\kappa^2_{-}\right)\,,
\end{equation}
where $\kappa^2_+ = - 2 (E + \sigma) \geq 0$ and $\kappa^2_- = 2
(E - \sigma) \geq 0$. $\kappa_g$ is then confined to the lie in
one of two intervals with a definite sign. It is thus confined to
inhabit a single hemisphere. It is intuitively clear that any such
state will spontaneously unbind into the interior of the sphere
where it may relax into a lower energy bound state.}

\vskip1pc \noindent The reconstruction of the loop from its
curvature data involves examining the conserved torque vector
${\bf M}$. Without loss of generality, it is always possible to
allign $\mathbf{M}$ along the $\hat{\mathbf{z}}$ axis, $\mathbf{M}
= M \hat{\mathbf{z}}$ (we follow Ref. \cite{JemalMartinBenAmar}). The normal
vector $\mathbf{n}$ is parametrized in terms of spherical polar coordinates
$\vartheta$ and $\varphi$,
\begin{equation}
 \mathbf{n}(s) = \left(\sin \vartheta (s) \cos \varphi (s),\sin \vartheta (s)
\sin \varphi(s) ,\cos
\vartheta (s)\right)\,.
\end{equation}
The projection of $\mathbf{M}$, given by Eq. (\ref{Torquevctsphcnf}),
onto ${\bf n}$ determines the polar angle in terms of
$\kappa_g$:\footnote{The projection onto $\mathbf{T} = \mathbf{n}'$ reproduces
(the derivative of) Eq. (\ref{Eq:theta}).}
\begin{equation} \label{Eq:theta}
\mathbf{M} \cdot \mathbf{n} = M \cos \vartheta = \kappa_g\,;
\end{equation}
its projection onto $\mathbf{l}$ determines the azimuthal angle $\varphi$
\begin{equation} \label{Eq:phi}
\mathbf{M} \cdot \mathbf{l} = - M\sin^2 \vartheta \varphi' = -
\left(\frac{\kappa^2_g}{2} +
\sigma - 1\right)\,.
\end{equation}
Thus the projections of $\mathbf{M}$
onto the Darboux frame determine the embedding functions
of the curve on the sphere in terms of $\kappa_g$ and the two constants,
$\sigma$ and $M$.

\vskip1pc\noindent
$\varphi$ will not generally increase monotonically along the loop. It will
exhibit overhangs with $\varphi'=0$, if $\sigma \le 1$. Combining
Eqs. (\ref{Eq:theta}) and (\ref{Eq:phi}) gives
\begin{equation} \label{Eq:phitheta}
\varphi' = \frac{M}{2}\, \left(\frac{ M^2 + 2 (\sigma - 1)}{ M^2 -
\kappa_g^2}-1\right)\,.
\end{equation}
Note that Eq. (\ref{Eq:theta}) implies the bound on $\kappa_g$,
$|\kappa_g| \le M$, already implicit in the quadrature,
Eq. (\ref{CEEsphereM2}).\footnote{A sharper, if less transparent,
bound follows from Eq. (\ref{CEEspherefirstint}) which implies, for
positive $\sigma$, $\kappa_g^2 \le 2 \sqrt{ M^2  + 2\sigma -1} - 2
\sigma$.} As we will see, $M$ will always be bounded. Thus
$\kappa_g$ and with it the geodesic energy will also be bounded.
Equation (\ref{Eq:theta})  also implies that the extremal values of
$\kappa_g$  occur where the polar angle is turning. The bound is
saturated when the loop passes through the poles.

\vskip1pc \noindent
Using the identification, (\ref{Eq:theta}),
it is possible to recast the quadrature in terms of $\theta$.  One has
\begin{equation}
\label{eq:quadtheta}
\theta'{}^2 + \frac{1}{M}\left( \frac{M^2 + 2(\sigma -1)}{\sin\theta} - M^2
\sin\theta \right)^2 = 1\,.
\end{equation}
In this form, it is clear that access to the poles is possible only when
$M$ and $\sigma$ are tuned so that the coefficient of the divergent term in the
potential appearing in Eq. (\ref{eq:quadtheta}) vanishes:
\begin{equation}
\label{eq:S}
M^2 + 2(\sigma -1)=0\,.
\end{equation}
This will require $\sigma < 1$.

\vskip1pc \noindent
When Eq. (\ref{eq:S}) is satisfied, the evolution of $\varphi$ simplifies.
Equation (\ref{Eq:phitheta}) assumes the form $\varphi'=- M/2$ so that
$\varphi$ increases linearly with arclength along the loop.  In general, the
dependence of $\varphi$ on $s$ is not monotonic, much less linear. This
identity, in turn, implies the value $M=2/R$ (and as a consequence
of Eq. (\ref{eq:S}), $\sigma=1 - 2/R^2$ at pole crossing.

\vskip1pc \noindent Let us first suppose that the loop is
sufficiently small so that neither self-contact nor
self-intersections occur. The periodic motion in the potential
$V(\kappa_g)$ implies an $n$-fold dihedral symmetry: the loop
closes upon completing $n$ periods of $\theta$ in one revolution of the polar
axis so that $\theta (s + 2\pi R/n)= \theta (s)$ and $\varphi(s+
2\pi R/n) = \varphi(s) + 2 \pi /n$, where $n\ge 2$ is an
integer.\footnote{The identity (\ref{Eq:theta}) implies that the
former is equivalent to $\kappa_g (s + 2\pi R/n)= \kappa_g (s)$.
The quadrature then implies that closure is smooth.} Closure
provides a quantization of physical states.\footnote { A one-fold
$n=1$ is incompatible with the four-vertex theorem for a sphere so
does not occur.}

\vskip1pc\noindent In equilibrium,
the number of circuits of the polar axis, $p$,
will increase with loop length, so that the boundary condition on
$\varphi$ is replaced with $\varphi(s+ 2\pi R/n) = \varphi(s) + 2 \pi p/n$.
Using Eqs. (\ref{Eq:theta}) and (\ref{Eq:phitheta}), and the quadrature
(\ref{CEEspherefirstint}) it is possible to cast the boundary conditions in the
form
\begin{equation}
\label{eq:c1}
\frac{2\pi  R}{4n}= \int_0^{k_1} \frac{d\kappa_g}{\sqrt{E^2 -  V(\kappa_g)}}\,,
\end{equation}
and
\begin{equation}
\label{eq:c2}
\frac{2\pi p}{4n} =
\frac{M}{2}\, \int_0^{k_1} \frac{d\kappa_g}{\sqrt{E^2 -  V(\kappa_g)}}\,
\frac{ \kappa_g^2 + 2 (\sigma - 1)}{ M^2 - \kappa_g^2}\,,
\end{equation}
where $k_1$ is the turning point of the potential, defined by
Eq. (\ref{eq:turningpoints}). Equation (\ref{eq:c1}) is independent of $p$,
whereas Eq. (\ref{eq:c2}) is independent of $R$. Together, they determine the
two constants of integration $\sigma$ and $M$ in terms of the loop radius $R$
and the two integers $n$ and $p$.

\section{Weak confinement by Spheres} \label{weak}

While Eq. (\ref{CEEsphereM2}) can be integrated exactly in terms of
elliptic functions \cite{LangSing, SingerSantiago, Arroyo}), a
perturbative approach to the problem is instructive.  Let us thus
suppose that $\Delta R := R-1 \ll 1$. Such a loop is sufficiently
small that neither self-contact nor self-intersections occur.
We thus expand the function $\kappa_g$ as well as the constants
$\sigma$ and $M$ in powers of $\epsilon =\sqrt{\Delta R}$, the
small dimensionless parameter in the problem:
\begin{equation}
\kappa_g = \kappa_1 + \kappa_3  + \cdots\,;\quad M= M_0+ M_2 + \cdots\,;\quad
\sigma= \sigma_0 + \sigma_2 + \cdots\,.
\end{equation}
The equilibrium states are described by small oscillations about
a geodesic circle on the sphere with $\kappa_g=0$.  The harmonic
approximation of the quadrature
(\ref{CEEsphereM2}) about $\kappa_g=0$ reads
\begin{equation} \label{linCEELsphere}
(\kappa_1')^2 + \sigma_0 \kappa_1^2  = M_0^2 - (\sigma_0-1) ^2 + 2 M_0 M_2 - 2
\sigma_0 \sigma_2\,.
\end{equation}
At lowest order, the arclength coincides with the azimuthal angle, $\varphi$,
and the geodesic curvature along a closed loop is given by
\begin{equation} \label{linear}
\kappa_1 (s)
\approx A_1\cos n \varphi \,,
\end{equation}
where $n$ is an integer and $A_1$ is a
constant.
$\Delta R$ is determined by the amplitude $A_1$ and the $n$ (see
\ref{bcdetail}).

\vskip1pc \noindent
The quadrature  implies that
$\sigma_0 = n^2$ and $M_0 = n^2 -1$. It also implies the
constraint
\begin{equation}
\label{eq:k1sM}
\frac{2}{n^2}(n^2-1) (M_2 - \sigma_2)= A_1^2
\end{equation}
on the difference of their second order corrections.

\vskip1pc \noindent
Equation (\ref{linear}) implements the boundary
conditions at lowest order. To complete the specification of
$\sigma_2$ and $M_2$ in terms of $\Delta R$, it is  necessary to
examine the boundary conditions (\ref{eq:c1}) and (\ref{eq:c2})
correct at next to leading order.  One finds that,  for $n\ge 2$,
Eqs. (\ref{eq:c1}) and (\ref{eq:c2}) together imply
\begin{equation} \label{eq:delR1}
\Delta R = - \frac {1}{4n^4} (\sigma_2 - M_2) - \frac{1}{4n^2} ( \sigma_2 + M_2)
-
\frac{1}{16 n^2}  A_1^2 \,,
\end{equation}
and
\begin{equation} \label{eq:delR2}
\Delta R= -
\frac{1}{n^2-1}( \sigma_2 - M_2) -  \frac{1}{4(n^2-1)}\frac{n^2+1}{n^2-1}\,
A_1^2\,.
\end{equation}
The details of the derivation are provided in \ref{bcdetail}. Using
Eq. (\ref{eq:k1sM}) in (\ref{eq:delR2}) one reproduces the relationship
Eq. (\ref{eq:A1delR}) between $A_1$ and $\Delta R$, $A_1^2 = 4 (n^2 -1) \Delta
R$.
It then follows that, for $n\ne 1$,
\begin{equation}
\label{eq:sig2}
\sigma_2 = \frac{1}{2} (3-7n^2) \Delta R\,,\quad M_2= \frac{3}{2} (1-n^2)\Delta
R\,.
\end{equation}

\subsection{Energy and transmitted force}

The bending energy of the  loop confined by the sphere
decomposes into a sum of geodesic
and normal parts,  reflecting the decomposition of the Frenet
curvature, $\kappa^2=\kappa_g^2 +1$,
\begin{equation} \label{Htotsphere}
 H = \frac{1}{2} \oint ds \left(\kappa^2_g + 1 \right) := H_g + H_n\,.
\end{equation}
$H_n = \pi R$ is the energy associated with an elastic rod that
been wound into a circular coil of radius $R_0=1$. It grows
linearly with loop length $2\pi R$ and is state independent. For a
weakly confined loop
\begin{equation}
\label{eq:Hweak}
H/H_{\sf loop}  \approx 1 + 2   n^2 \Delta R \,,
\end{equation}
where $H_{\sf loop} = \pi/R$
is the bending energy of a circular loop of radius $R$. The energy
increases linearly with loop size. This will not be true in longer loops.

\vskip1pc \noindent For a fixed value of $\Delta R$, the energy
increases quadratically with $n$. The ground state, as one would
have predicted, is the completely attached $n=2$ state. As we show, all states
with $n\ge 3$ are unstable.

\vskip1pc \noindent
The force transmitted to the sphere at any point, given in Eq.
(\ref{CEELmagnlambda}), takes the particularly simple form
\begin{equation} \label{lambdasphere}
\lambda = \frac{1}{2} \, \kappa^2_g + \sigma \,.
\end{equation}
It differs from the local energy density: compare Eqs. (\ref{Htotsphere}) and
(\ref{lambdasphere}). Its spatial dependence is completely determined by the
geodesic curvature. It is bounded from below by $\sigma$. This lends a direct
physical interpretation of $\sigma$. For small $\Delta R$ and each $n\ge 2$, 
the force is given  by
\begin{equation}
\label{eq:Fweak}
\lambda_n \approx n^2 + \frac{\Delta R}{2} \left[ 3 - 7 n^2 +
4 \left(n^2-1\right) \cos^2 \, n \varphi \right]\,;
\end{equation}
while it oscillates along the loop, it remains positive everywhere.
$\lambda_n$  does not vanish in the limit $\Delta R\to 0$.  This is
interpreted as an Euler instability. The buckling of the circular
loop into an $n$-fold involves a critical compression in the loop
provided by this normal force.  Thereafter, $\lambda_n$ decreases
linearly with $\Delta R$. This may appear counterintuitive. This
behavior will be put in context when we examine loops outside of
perturbation theory.

\subsection{Stability}

Here we examine the stability of the weakly confined $n$-folds that have been
described. To do this we require the second variation of the energy with respect
to small deformations of the loop. This is expressed in the form
\cite{Stability}
\begin{equation}
\delta^2  H= \int d\varphi\, \Phi{\cal L}\Phi\,,
\end{equation}
where  $\Phi$ is the deformation along ${\bf l}$, and the self-adjoint operator
${\cal L}$ is given by
\begin{equation}
{\cal L}= \frac{\partial^4}{\partial \varphi^4} + (n^2+1) \,
\frac{\partial^2}{\partial \varphi^2}  + n^2\,.
\end{equation}
The fixed length constraint along the spherical surface implies a
global constraint on the normal deformation,
\begin{equation} \label{eq:arccon1}
\oint ds \kappa_g \Phi= 0\,.
\end{equation}
Periodicity  implies that the modes of deformation are represented
by a constant ($m=0$),  $\sin m \varphi$ and $\cos m \varphi$,
$m=1,2,3,\dots$. For fixed $n$, the eigenvalues of ${\cal L}$ are
then labeled by the integer $m=0,1,2,\dots$, given by
\begin{equation}
C_m = (m^2- n^2)( m^2-1)\,.
\end{equation}
The constant mode with $m=0$ has positive $C_0= n^2$.
All eigenvalues with $m\ge 1$ possess a two-fold degeneracy.

\vskip1pc\noindent There are four zero modes satisfying ${\cal L}
\Phi=0$; two occur at $m=1$ and two at $m= n$. The mode $\sin n
\varphi \propto \kappa_g'$ corresponds to rotation of the loop
about the axis of symmetry.  The two modes with $m= 1$ correspond
to rotations about an orthogonal axis. These three modes are the zero modes
anticipated by the rotational invariance of the bending energy. The fourth zero
mode $\cos n\varphi$ is inconsistent with the fixed length constraint
(\ref{eq:arccon1}) and, so, is unphysical. It is also the only mode of
deformation inconsistent with this constraint in this regime.

\vskip1pc\noindent
The two-fold ground state with $n=2$ is stable. There are no modes of
deformation
with negative eigenvalue.

\vskip1pc\noindent All excited confined states are unstable. There
will be $2(n-2)$ unstable modes of deformation corresponding to
$m=2, \dots, n -1$ lying between the zero modes at $m=1$ and
$m=n$. All modes of deformations with $m > n$ contribute a
positive energy. The first excited state with $n=3$ has two modes
of decay, of equal energy, into the ground state. The dominant
mode of instability in higher energy states is not directly
towards the ground state involving a cascade of instabilities.

\section{Strong spherical confinement} \label{Strong}

Let us now examine the shape adopted by a loop of finite $R$ confined by the
sphere.

\vskip1pc \noindent
Equation (\ref{CEELsphere}) can be integrated in terms of elliptic
functions to give
$\kappa_g$ as a function of $s$ \cite{LangSing, SingerSantiago, Arroyo}
\begin{equation} \label{Eq:kappagab}
\kappa_g (s) = \kappa_0 \, \text{cn} \,\left[q s,m\right]\,, \qquad \kappa_0 = 2
\sqrt{m} q\,.
\end{equation}
The function $\text{cn}[x,m]$ is the Jacobi elliptic cosine
\cite{AbramStegun}. The angular wavenumber $q$ is given in terms
of the constant $E$ defined below Eq. (\ref{eq:Vdef}) by
$q=\sqrt{E}$; the modulus $m$ is defined by
\begin{equation}
m =\frac{1}{2}\left(1-\frac{\sigma}{q^2}\right)\,.
\end{equation}
The curvature depends on the two parameters $\sigma$ and $M$
through the parameters $q$ and
$m$.\footnote {Definitions of $q$ and $M$ are inverted to give
\begin{equation} \label{Eq:CEEsigmaM}
 \sigma = q^2\, \left(1-2 m \right) \,, \qquad M^2 =
\left(q^2 -1\right)^2 + 4 m q^2\,.
\end{equation}
The modulus $m$ will lie in the interval $[0,1]$; this bounds
$\sigma$ in terms of $q$: $|\sigma| \le  q^2$, changing sign when
$m = 1/2$.} These parameters will be determined explicitly in
terms of $R$ and $n$ using the boundary conditions associated with
the closure of the loop. \vskip1pc \noindent Using the fact that
the period of $\text{cn}$ is given by $4 \mathcal{K}[m]$, where
$\mathcal{K}[m]$  is the complete elliptic integral of the first
kind \cite{AbramStegun}, it is possible to cast the boundary
condition on $\kappa_g$, given by Eq. (\ref{eq:c1}), in the form
\begin{equation}
\label{eq:bck}
q = 4 n\, \mathcal{K}[m]/(2\pi R)\,.
\end{equation}
Integration of Eq. (\ref{Eq:phi}) gives
\begin{equation}
\varphi(s) = \frac{M}{2} \left(\frac{q^2+ 1}{q \left(q^2-1
\right)} \, \Pi \left[-\frac{4 m q^2}{(q^2 - 1)^2},\,
\text{am}\left[q \,s ,m\right],\, m\right] - s\right)\,,
\end{equation}
where $\Pi[\eta,\text{am}[x,m],m]$ is the incomplete elliptic integral of the
third kind and
$\text{am}\left[x,m\right]$ is the Jacobi amplitude \cite{AbramStegun}.
Closure of the loop after $p$ circuits of the polar axis, $\varphi(2\pi R) = 2
\pi p$, then reads
\begin{equation}
\label{eq:varphiSm}
\frac{2\pi M \, R}{2} \left(\frac{q^2 + 1}{{\cal K}[m] \left(q^2 -1\right)} \,
\Pi
\left[-\frac{4 m q^2}{\left(q^2 - 1\right)^2},\, m\right] - 1\right) =
2\pi p\,.
\end{equation}
Modulo Eq. (\ref{eq:bck}), Eq. (\ref{eq:varphiSm}) determines
$m$ implicitly as a function of  $R$. This completes the formal
construction of the confined loops.

\subsection{Ground state $n=2$, $p=1$ and its descendants}

The two-fold state with $n=2$ is illustrated in Fig. \ref{figure1}
for values of $R$ in the interval $[1,3]$. When  $\Delta R$ is
small, it consists of a small oscillation about a geodesic circle
(see Fig. 1(a)) consistent with the perturbative description. As
$R$ is increased the oscillations about this circle initially
increase in amplitude as more of the surface of the sphere is
explored (Figs. 1(b) and 1(c)).  In Fig. 1 (c) overhangs with
$\varphi'=0$ occur on the increasingly crowded loop. At a critical
value $R_2= 2.127$, the loop makes self-contacts at the two poles
(Fig. 1(d)).\footnote {For each $n$ there will be a critical
value $R_n$ where this occurs, whose magnitude increases with $n$.
As pointed out in \cite{JemalMartinBenAmar}, this value saturates.
Beyond some critical loop length, all equilibrium states are
self-intersecting.}

\vskip1pc \noindent For values of $R$ above $R_2$, one needs to look
more carefully at the boundary conditions. If, as we assume,
the microscopic physics accommodates self-intersections on the
surface and they occur without costing energy, the mathematical
curve described by the elliptic function continues to represent the
physical loop.\footnote{If self-intersection is prohibited, the
physics will be very different. It has been described in the conical
context in Ref. \cite{JemalMartinBenAmar}, and explored
numerically  in detail in Ref. \cite{Stoop2010}. As the loop
crowds itself on the sphere, there will be a steep rise in the
energy associated with this self-confinement. Beyond some point,
partially attached loop conformations would be expected to become
energetically favored.}

\vskip1pc \noindent There is a qualitative change in the behavior as
$R$ is increased through $R_2$. On crossing the poles,  the loop
makes two additional revolutions about the polar axis so that $p=1$
must be replaced by $p=3$ in the boundary condition
Eq. (\ref{eq:c2}). To see this, note that the loop will cross the
poles four times, twice at each pole. A discontinuity of $-2\pi$ is
introduced in $\varphi$ at each crossing. These discontinuities
contribute to its period; thus, the period of $2\pi$ describing a
single revolution gets replaced by $2\pi- 4 (2\pi) = - 6\pi$. The
orientation is also reversed.\footnote{ For an $n$-fold, the
corresponding number of revolutions is $p=2n-1$.} As Fig.
\ref{figure1} illustrates, this discontinuity marks a transition
from oscillatory to orbital behavior.

\vskip1pc\noindent As $R$ is increased further the two regions
bounded by the self-intersections grow (Fig. 1(e)). When $R=3$ the
loop degenerates into a triple cover of a geodesic circle (Fig.
1(f)).

\subsubsection{States as trajectories in parameter space}

\vskip1pc \noindent Each equilibrium state can be represented as a
point on the parameter space $(\sigma,M)$.  As $R$ increases, the
state will follow a trajectory  $R\to (\sigma(R),M(R))$ labeled
by the two integers $n$ and $p$.\footnote {The boundary condition
on $\varphi$, Eq. (\ref{eq:c2}), does not involve $R$ explicitly.
It thus identifies the trajectories on the $(\sigma, M)$ parameter
space without locating the position along these trajectories. The
latter is provided by Eq. (\ref{eq:c1}).}

\begin{figure}
\begin{center}
  \subfigure[Ground States $n=2$, $p=1$ and
descendants]{\includegraphics[scale=0.35]{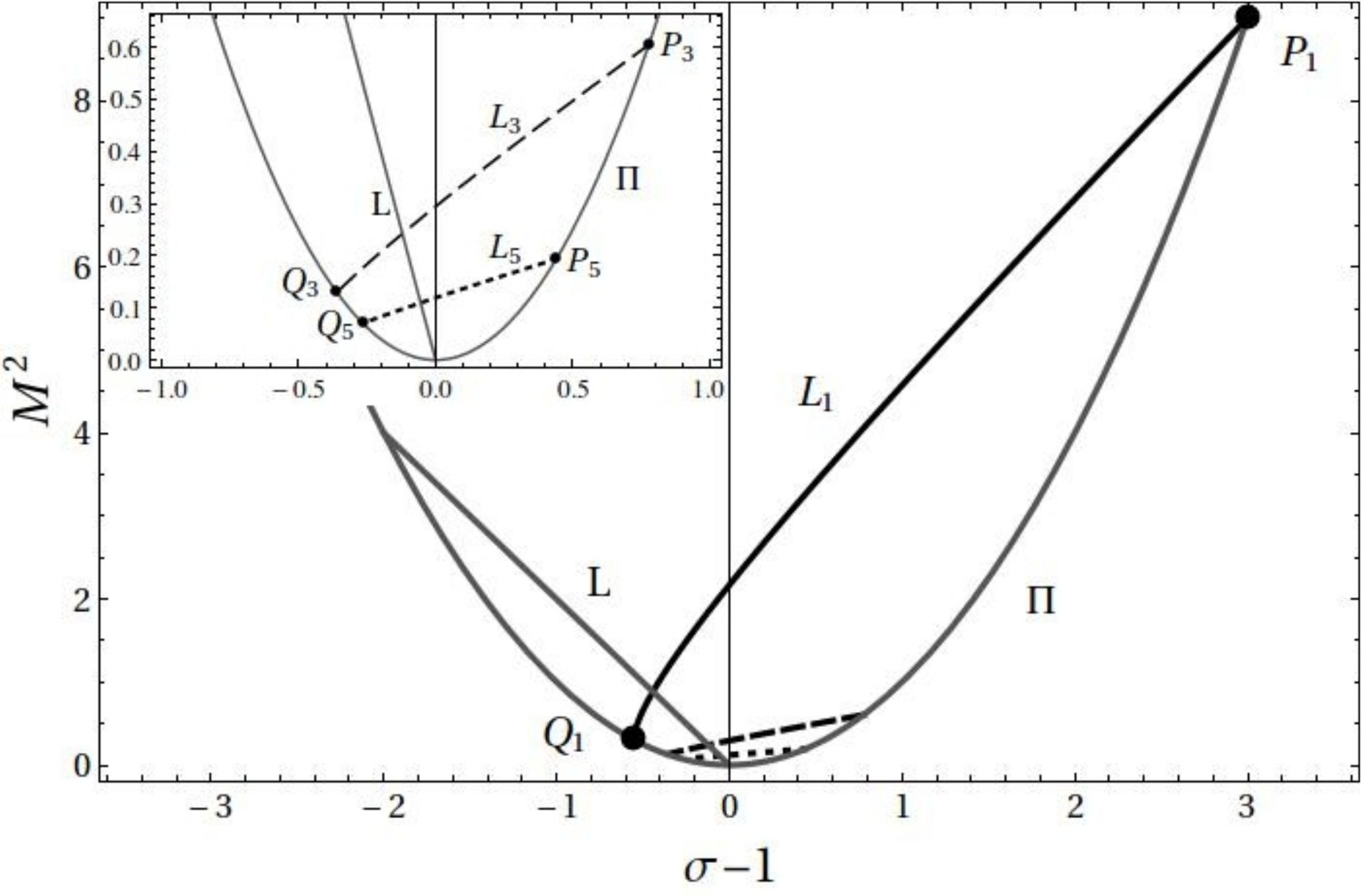}}
  \hfill
  \subfigure[Ground States $n=3$, $p=2$ and
descendants]{\includegraphics[scale=0.36]{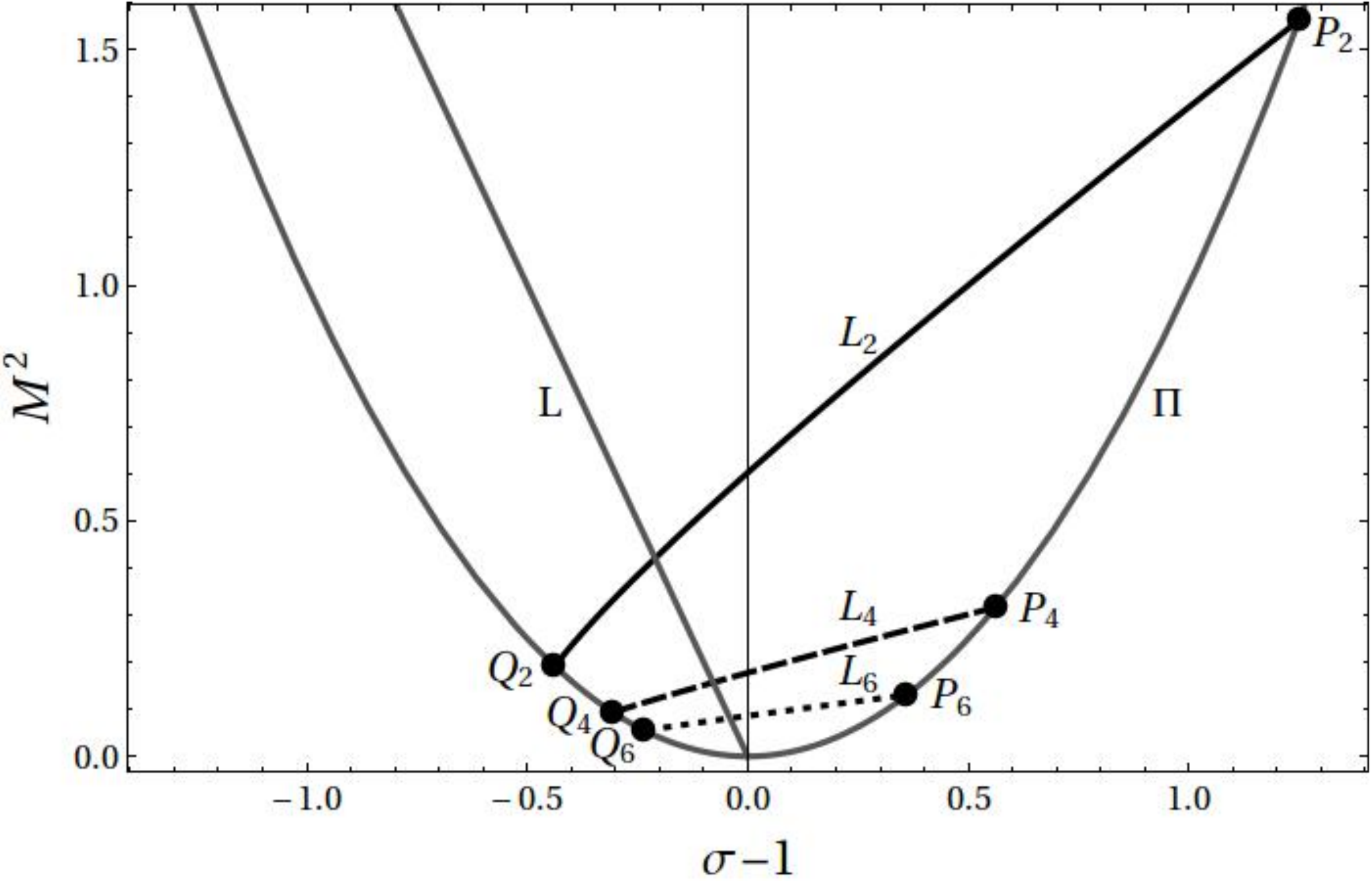}}
\end{center}
\caption{\small Trajectories in parameter space ($\sigma -1, M^2$).  (a) States
with $n=2$ are represented by
the solid curve $L_1$ for $R$ in the interval $[1,3]$. This curve originates at
the point $P_1=(3,9)$ when $R=1$ (geodesic circle)  and terminates at the point
$Q_{1}=(-5/9, 25/81)$ when $R_3$ (a triply covered geodesic circle).
Both lie on the boundary parabola $\Pi$. The straight line L indicates loops
passing through the poles.  The descendant states with $n=4$ ($n=6$) which occur
for $R$ in the interval $[3,5]$ ($[5,7]$)
are represented by the dashed curve $L_3$ (dotted curve $L_5$) in (a), where
they are also shown with zoom in the inset. The counterpart of (a) for the state
$n=3$, $p=2$ and its descendants is represented in (b).
States with $n=3$ for $R$ in the interval $[2,4]$, as well as the descendants
with $n=5$ and $n=7$ for $R$ in the intervals $[4,6]$ and  $[6,8]$ respectively,
are represented by the solid, dashed, and dotted curves, $L_2$,$L_4$,
$L_6$.}\label{figure2}
\end{figure}

\vskip1pc \noindent
To facilitate the interpretation of trajectories in parameter space, it
is useful to first locate relevant geometrical landmarks.  Various significant
parameter curves are represented in Fig. \ref{figure2}.

\vskip1pc \noindent (1) The parabola $\Pi$ given by $ M^2 =
(\sigma-1)^2$ (saturating the inequality (\ref{eq:parabola}))
provides a boundary on  parameter space.\footnote{For graphical
purposes we plot $M^2$ vs $\sigma$; hence ``parabola''.} Points on
the boundary describe geodesic loops, with $\kappa_g =0$ and
$\kappa_g'=0$, covering the equator an integer number of times.
Points below the parabola do not describe confined loops.

\vskip1pc \noindent (2)
Points on the straight line L, given by
$M^2 = 2 (1-\sigma)$,  describe loops that
pass through the poles.  Points above this line represent loops that display
oscillatory behavior; those below it display orbital behavior.

\vskip1pc \noindent (3)
Overhangs occur to the left of the vertical line, $\sigma=1$.

\vskip1pc \noindent The solid curve, $L_1$ in Fig. \ref{figure2}
represents the trajectory of the state with $n=2$ and $p=1$ as
$R$ is increased in the interval $[1,3]$. The initial state with
$R=1$--a single geodesic circle--is represented by the point
$P_1=(4,3)$ lying on the boundary parabola $\Pi$. In this
interval, both $\sigma$ and $M$ decrease monotonically with $R$.
The trajectory terminates in the point  $Q_1=(4/9, 5/9)$ on the
boundary parabola, where the loop degenerates into a triple
covering of a geodesic circle.\footnote {Prior to crossing L,  the
trajectory  $L_1$ is approximated accurately by the straight line,
$M^2= 9\left( 1+ 6/25\, (\sigma - 4)\right)$, with a slope
$54/25$,  predicted in perturbation theory. Perturbation theory
thus provides an accurate description of the problem outside its
expected range of validity.}

\begin{figure}
\begin{center}
\subfigure[$R = 3.05$]{\includegraphics[scale=0.12]{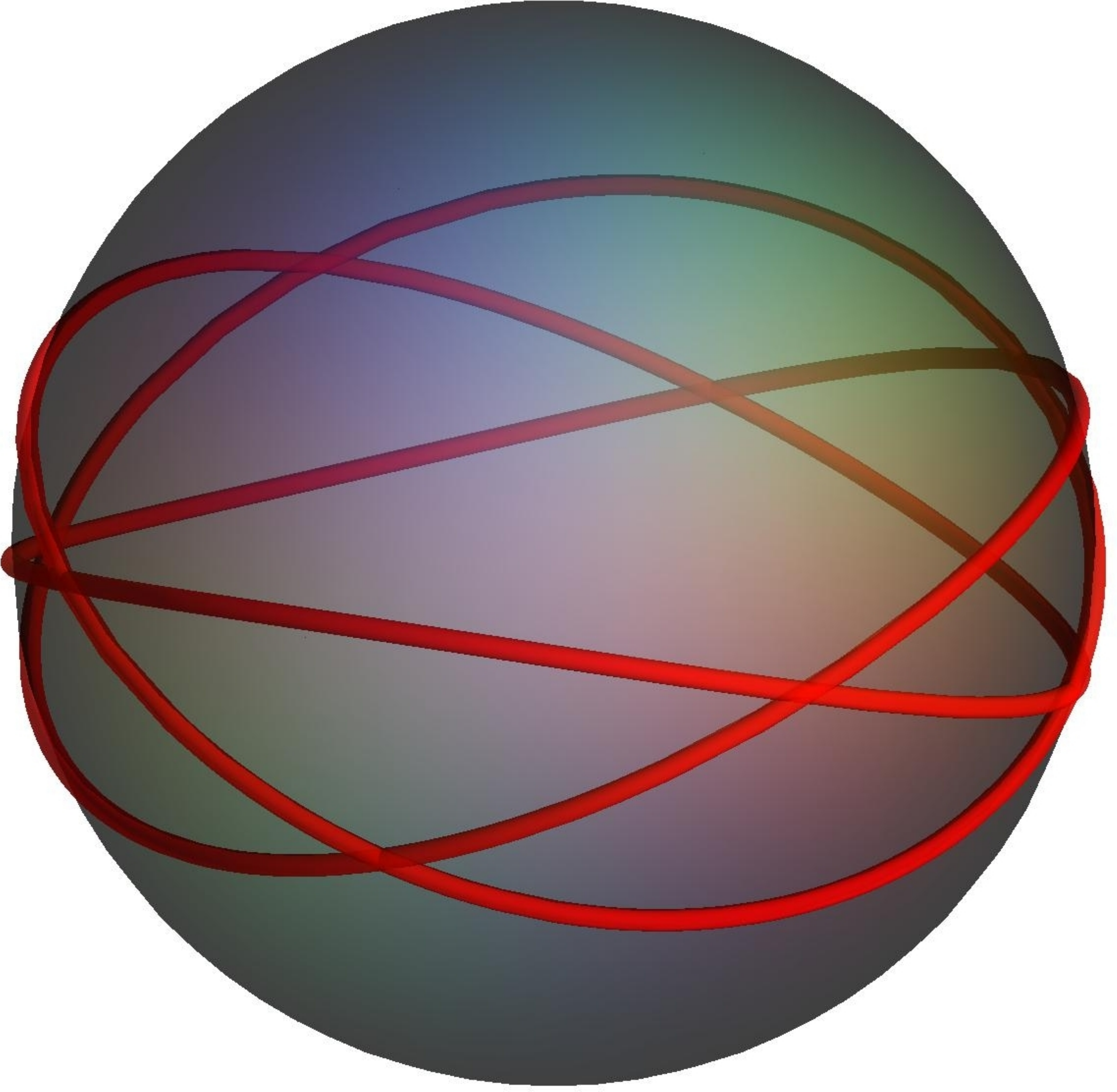}}
\hfill
\subfigure[$ R = 4.06$]{\includegraphics[scale=0.12]{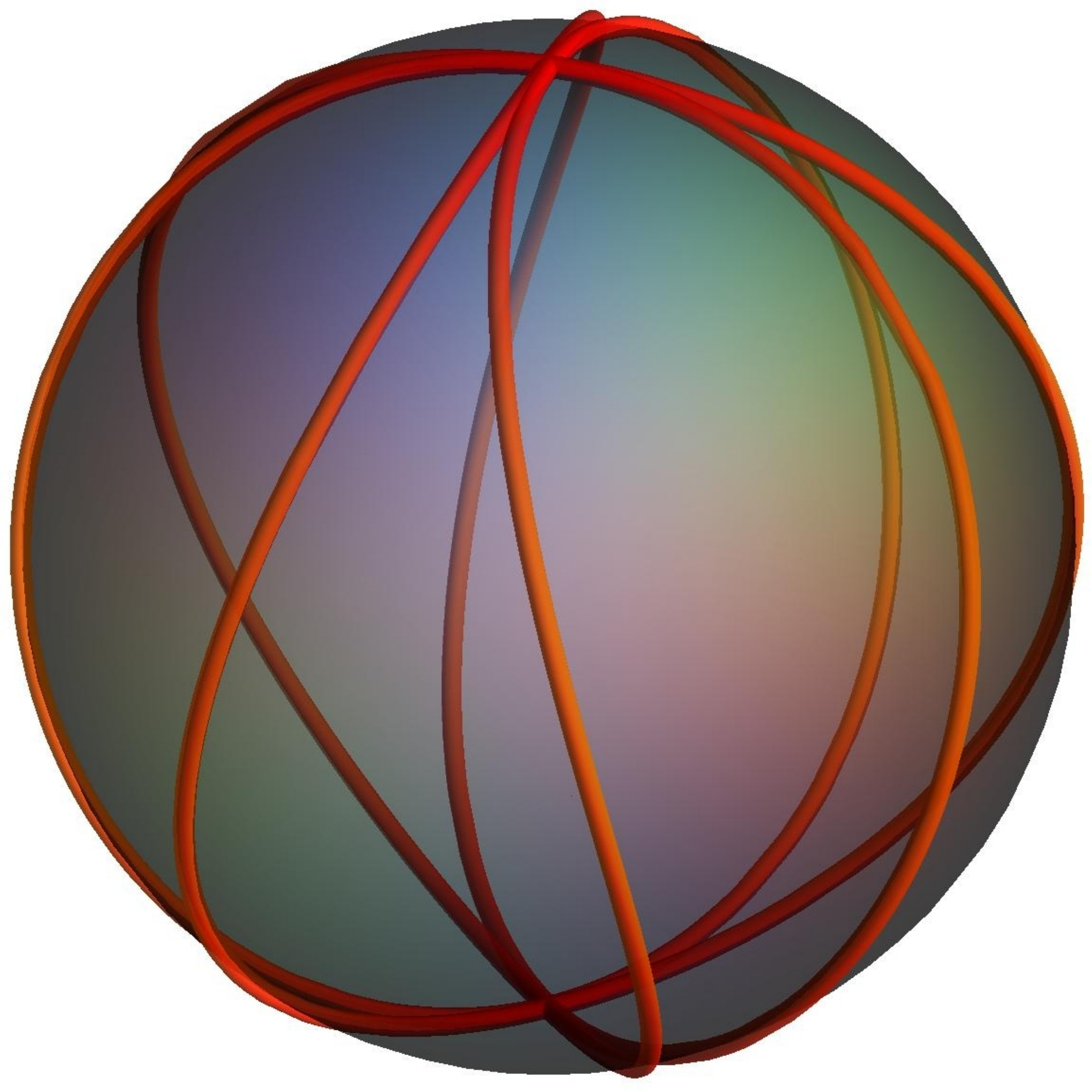}}
\hfill
\subfigure[$ R = 4.99$]{\includegraphics[scale=0.12]{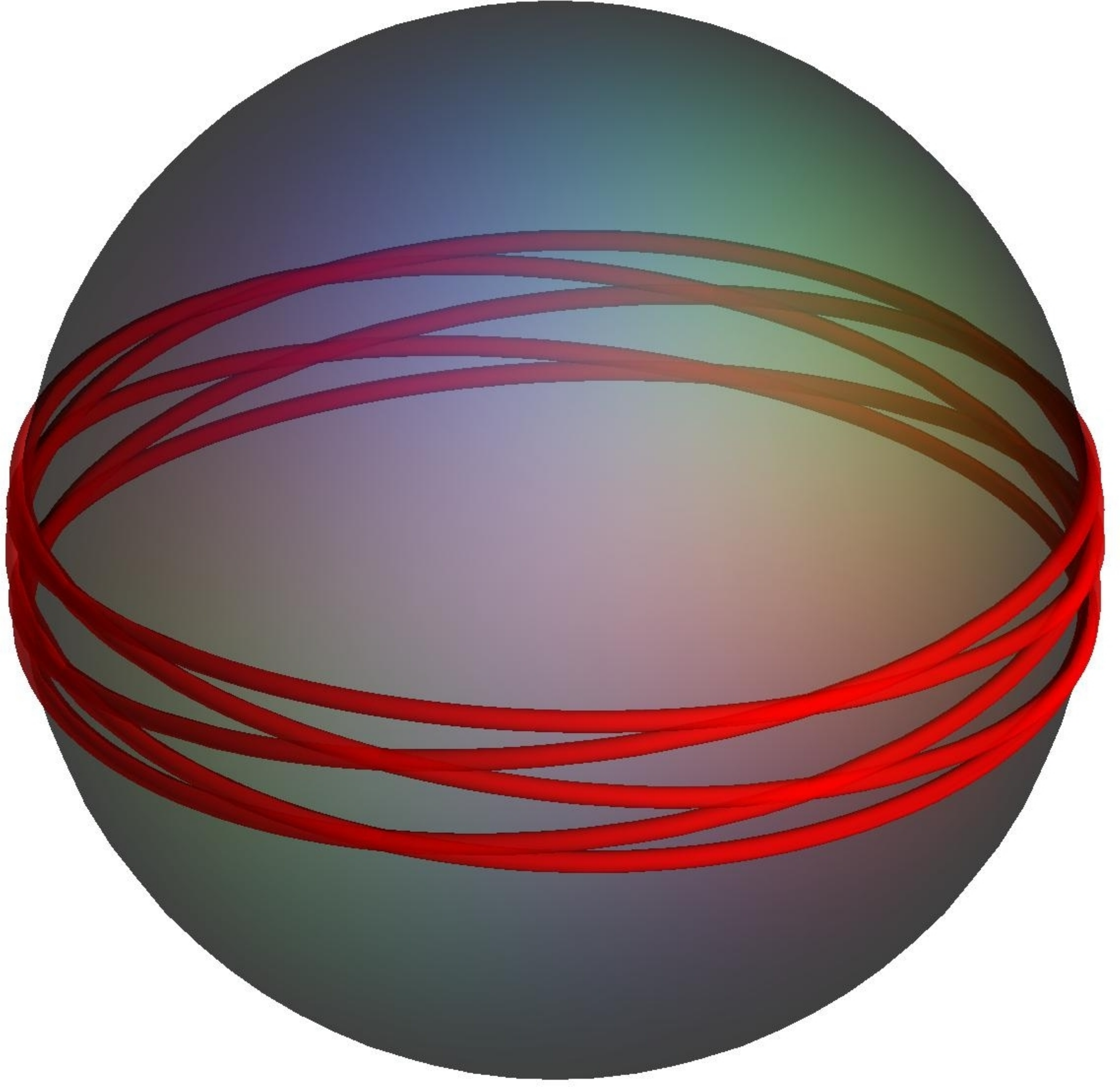}}\\
\includegraphics[scale=0.175]{fig1g.pdf}
\end{center}
\caption{\small (Color online) States with four-fold symmetry, $n=4$, $p=3$ for
$R$ in the interval $[3,5]$:  (a) increasingly deformed triple covering of the
equator;  (b) pole crossing, four points at a time); (c) state making five
orbits of sphere before it becomes a quintuple covering of equator. As in Fig. 1
colors represent the normalized local confining force.
}\label{figure3}
\end{figure}

\vskip1pc \noindent More generally, whenever $R=p$, where $p$ is
an integer, loops may form $p$-fold coverings of a geodesic
circle. As $R$ is raised above $R=p$, it is possible to examine
the state perturbatively in a manner analogous to that of small
loops. Suppose that the loop oscillates with an $n$-fold symmetry,
so that $\kappa_g \approx \cos n \varphi/p$.  Consistency with
$R>p$ places a lower bound on $n$: $n>p$. Thus one must have $n=
p+1, p+2,\dots$.\footnote{Note that the analog of Eq. (\ref{skappa}) for $p$
revolutions implies
\begin{equation} \label{skappap}
R/ p   \approx 1 + \frac{1}{4} \frac{k_1^2}{M_0^2} \left(\frac {n^2}{p^2} -1
\right) \,.
\end{equation}
} The state with $p+1$-fold dihedral symmetry will minimize the
energy among these states. It is thus identified as the ground
state. In particular, when $R$ passes through $R=3$, the symmetry
of the ground state changes from $n=2$ to $n=4$ (see Fig. \ref{figure3}).

\vskip1pc \noindent It is straightforward to determine the values
of $\sigma$ and $M$ in the ground state with $n=p+1$ as $R$ passes
through $R=p$: $\sigma (p) =(p+ 1)^2/p^2$, and $M (p)=(2p +
1)/p^2$, representing  a point lying on the boundary parabola
$\Pi$.

\vskip1pc \noindent As $R$ is increased the amplitude of
oscillation increases.  At some point the loop will (re)cross the
poles where the number of revolutions changes by two. As $R=p+2$ is approached,
the loop morphs continuously into a $p+2$ cover of a geodesic circle. In this
region, the ground state can also be treated perturbatively. Now $\kappa_g
\approx \cos ((p+1)/(p+2)  \varphi)$. The corresponding end point in parameter
space as $R$ approaches $R=p+2$ is given by $\sigma=(p + 1)^2/(p+2)^2$, $M=(2p
+3)/(p+2)^2$,  which also lies on the boundary parabola $\Pi$.

\vskip1pc \noindent The trajectory in the parameter plane
representing the $n=2$, $p=1$ ground state loop and its minimum
energy descendants is illustrated in Fig. \ref{figure2}. It
consists of a number of disconnected curve segments, $L_p$,
$p=1,3,5,\dots$ with end points $P_p=((p + 1)^2/p^2,(2p+1 )/p^2)$,
and $Q_p=((p + 1)^2/(p+2)^2,(2p +3)/(p+2)^2)$ lying on the
boundary parabola $\Pi$. The points $Q_p$ and $P_{p+2}$ describe
loops with $R=p$ as it is approached from below and above
respectively.

\vskip1pc \noindent There are discontinuities in $\sigma$ and
$M^2$ at $R=p=3,5,7,\dots$ and the loop passes through a $p$-fold
covering of a geodesic circle.  These discontinuities are
identified analytically using perturbation theory. Their origin is
the transition from orbital back to oscillatory behavior with a
change from $p-1$-fold to $p+1$-fold symmetry.

\vskip1pc \noindent
As  $p\to\infty$, $P_{p},Q_{p}\to (1,0)$ so that $\sigma\to 1$ and $M\to 0$.

\begin{figure}
\begin{center}
  \subfigure[]{\includegraphics[scale=0.35]{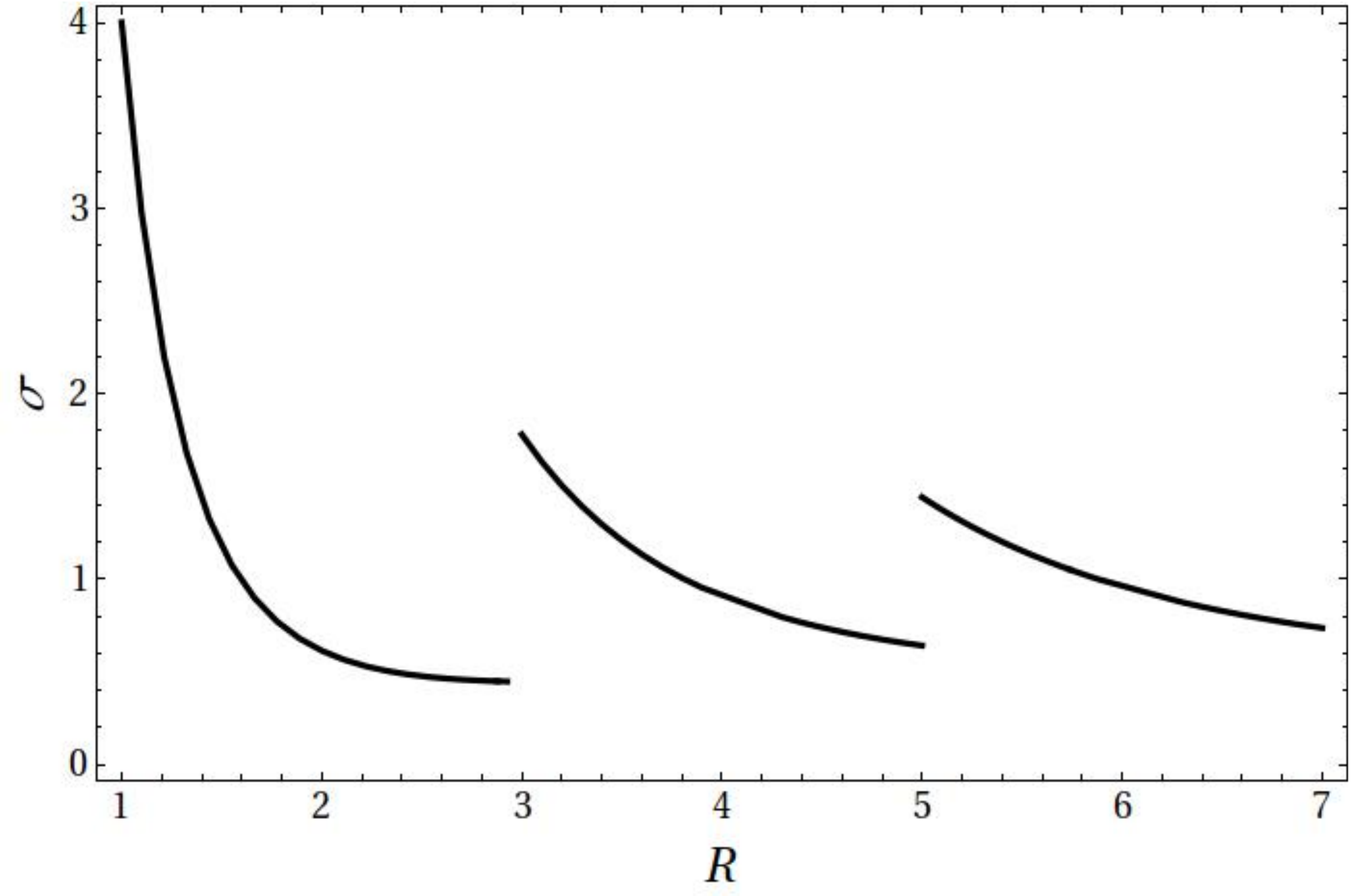}}
  \hfill
  \subfigure[]{\includegraphics[scale=0.36]{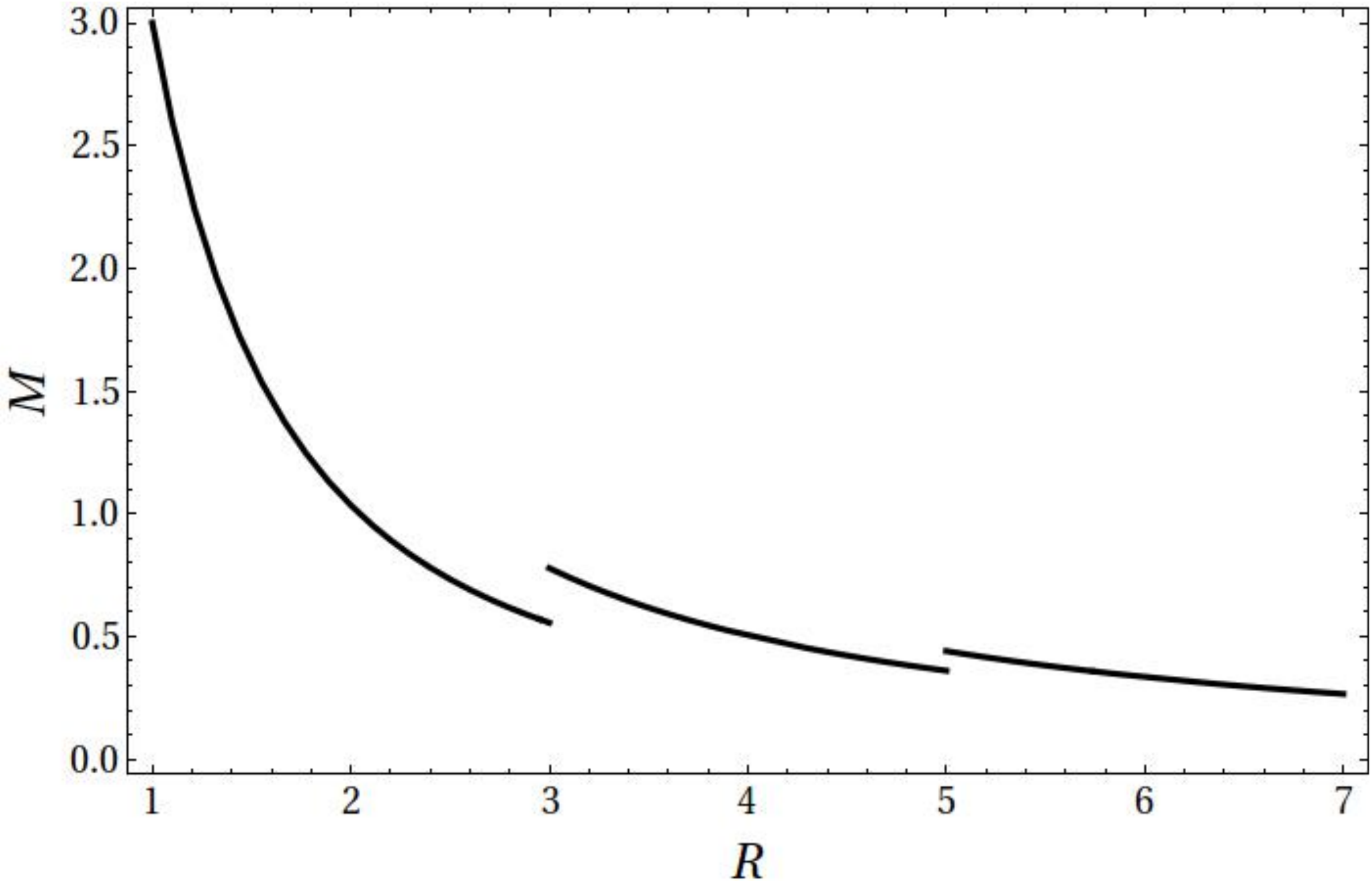}}
\end{center}
\caption{\small $\sigma$ (a) and $M$ (b)  vs. $R$ for $n=2,p=1$
and its descendants in the interval $[1,7]$, as described in the
text represented by the trajectories $L_1,L_3$ and $L_5$ in Fig.
\ref{figure2}.
 Asymptotically,  $\sigma \to 1$ and
$M\to 0$.}\label{figure4}
\end{figure}

\vskip1pc\noindent
In Fig. \ref{figure4} $\sigma$ and $M^2$ are plotted as functions
of $R$.

\subsection{Ground state $n=3$, $p=2$ and its descendants}

\begin{figure}
\begin{center}
  \subfigure[$R = 2.05$]{\includegraphics[scale=0.12]{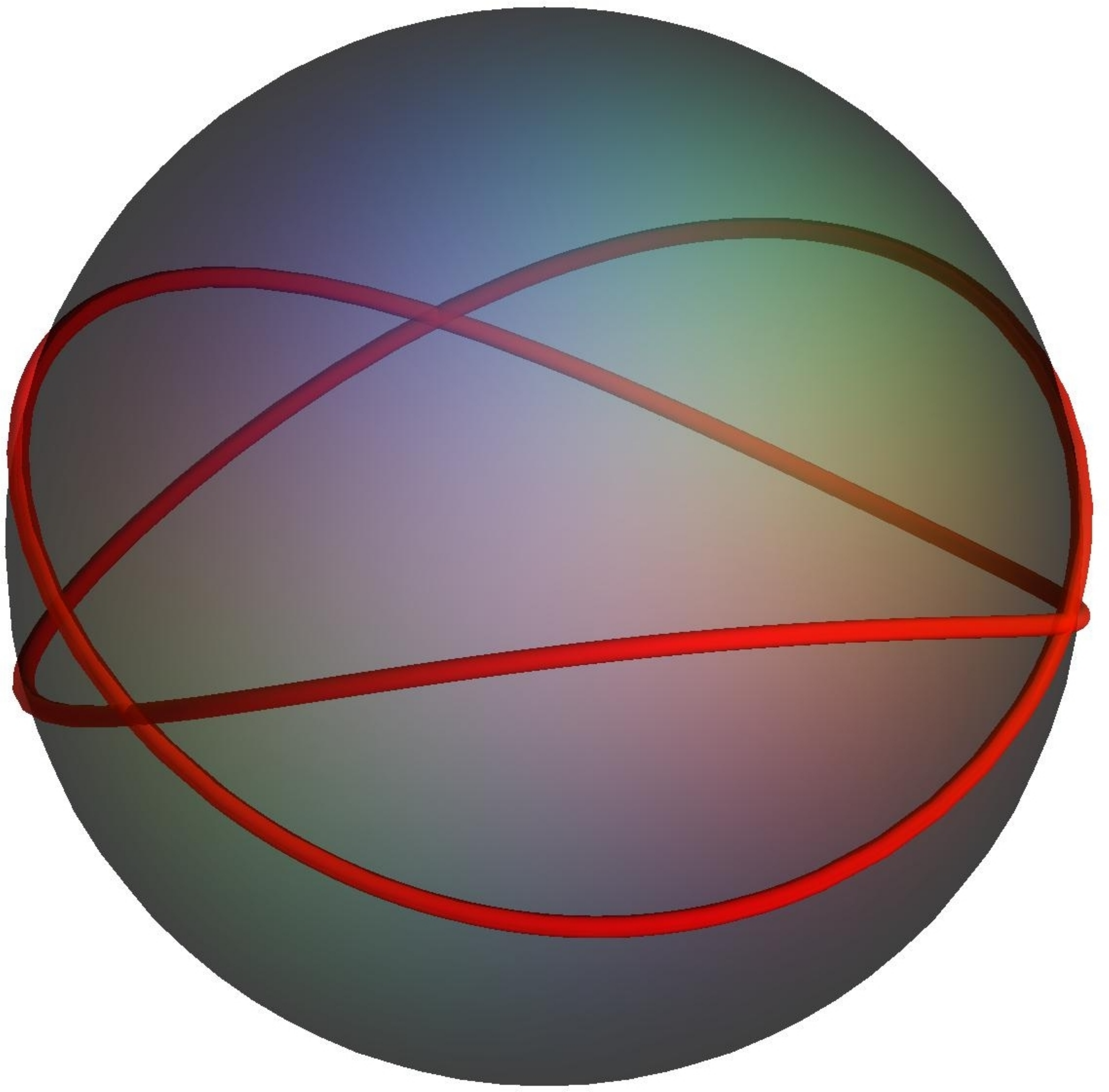}}
  \hfill
  \subfigure[$ R = 2.5$]{\includegraphics[scale=0.12]{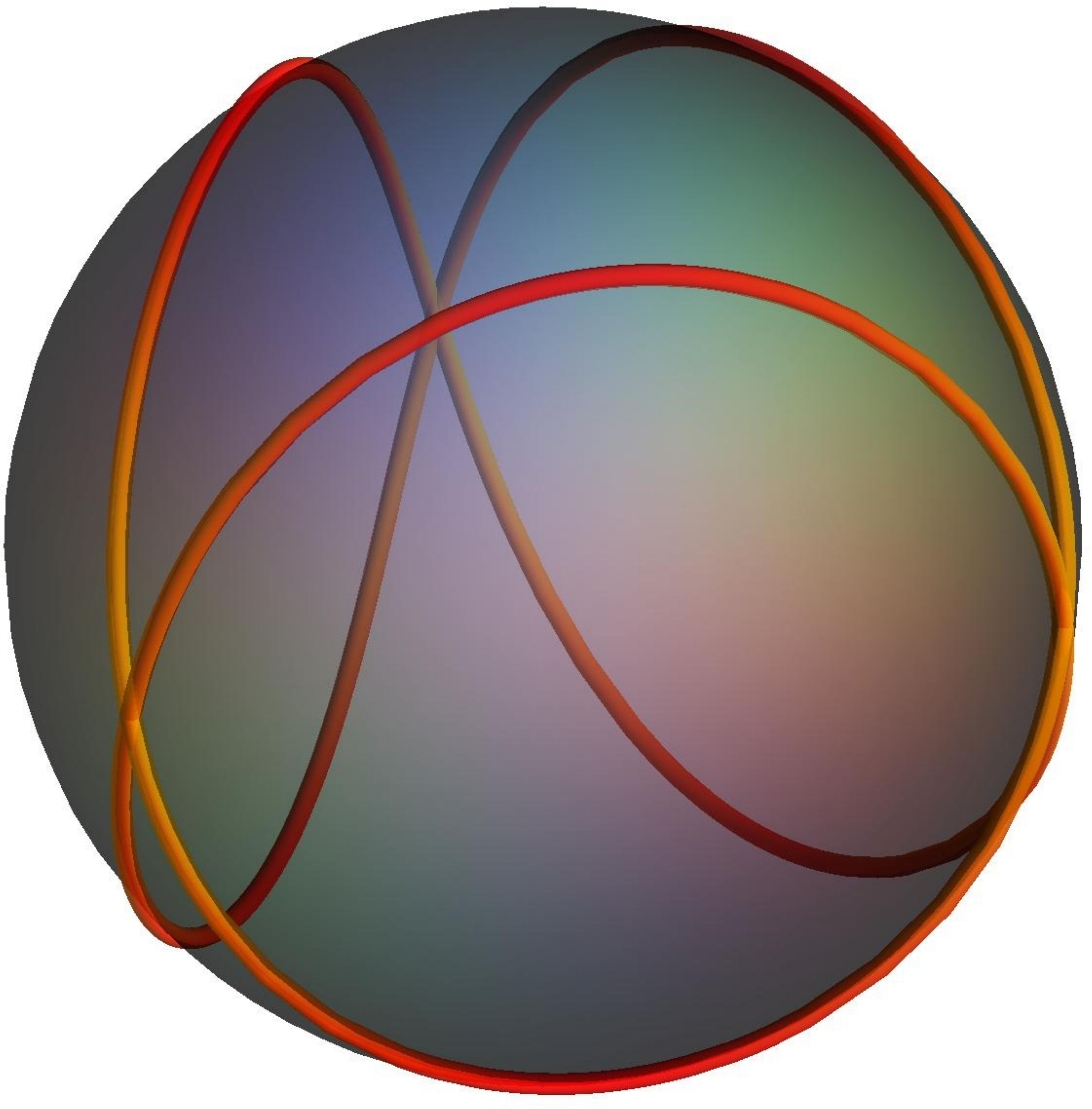}}
  \hfill
  \subfigure[$ R = 3.0$]{\includegraphics[scale=0.12]{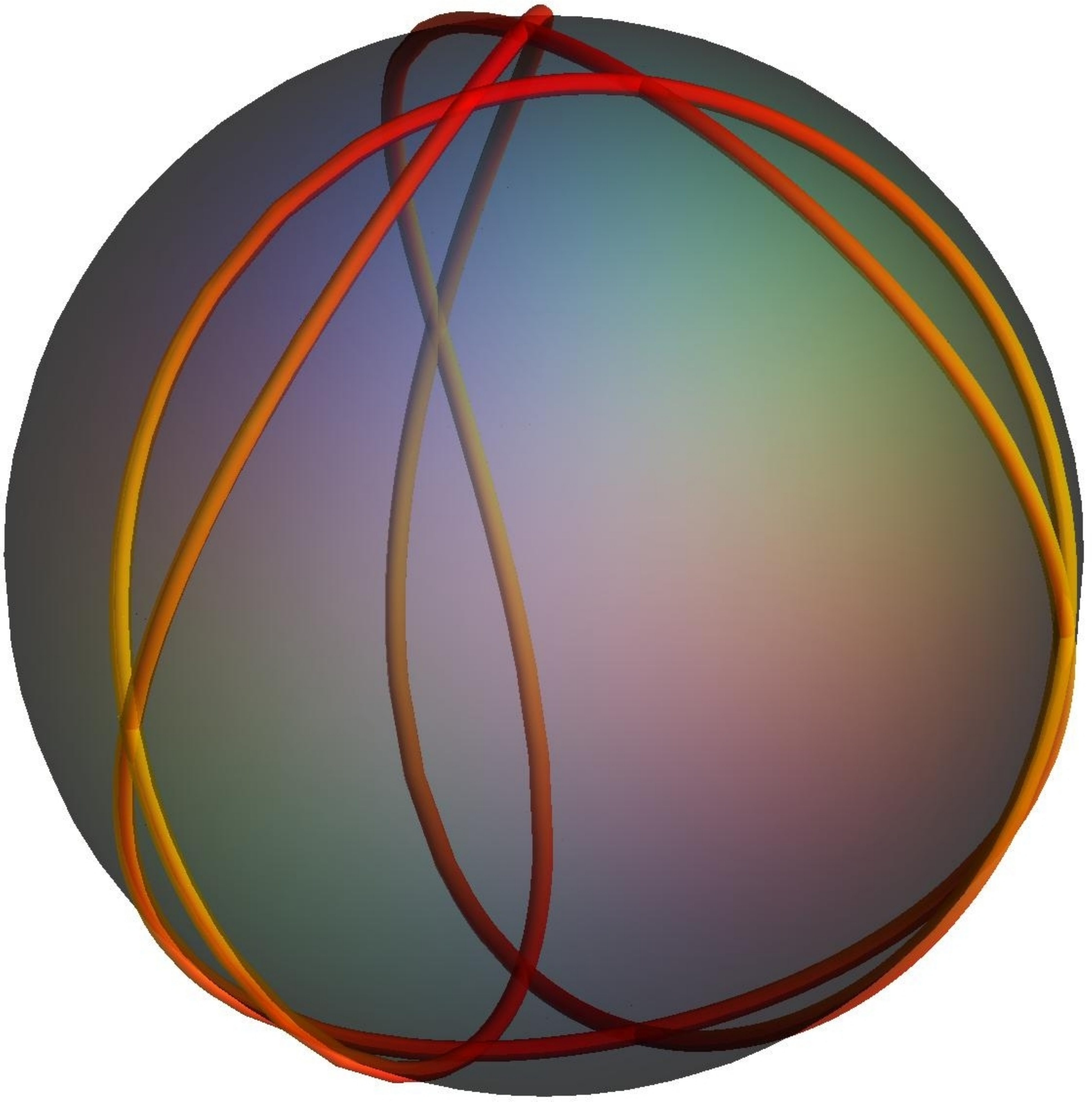}}\\
  \subfigure[$ R = 3.08$]{\includegraphics[scale=0.12]{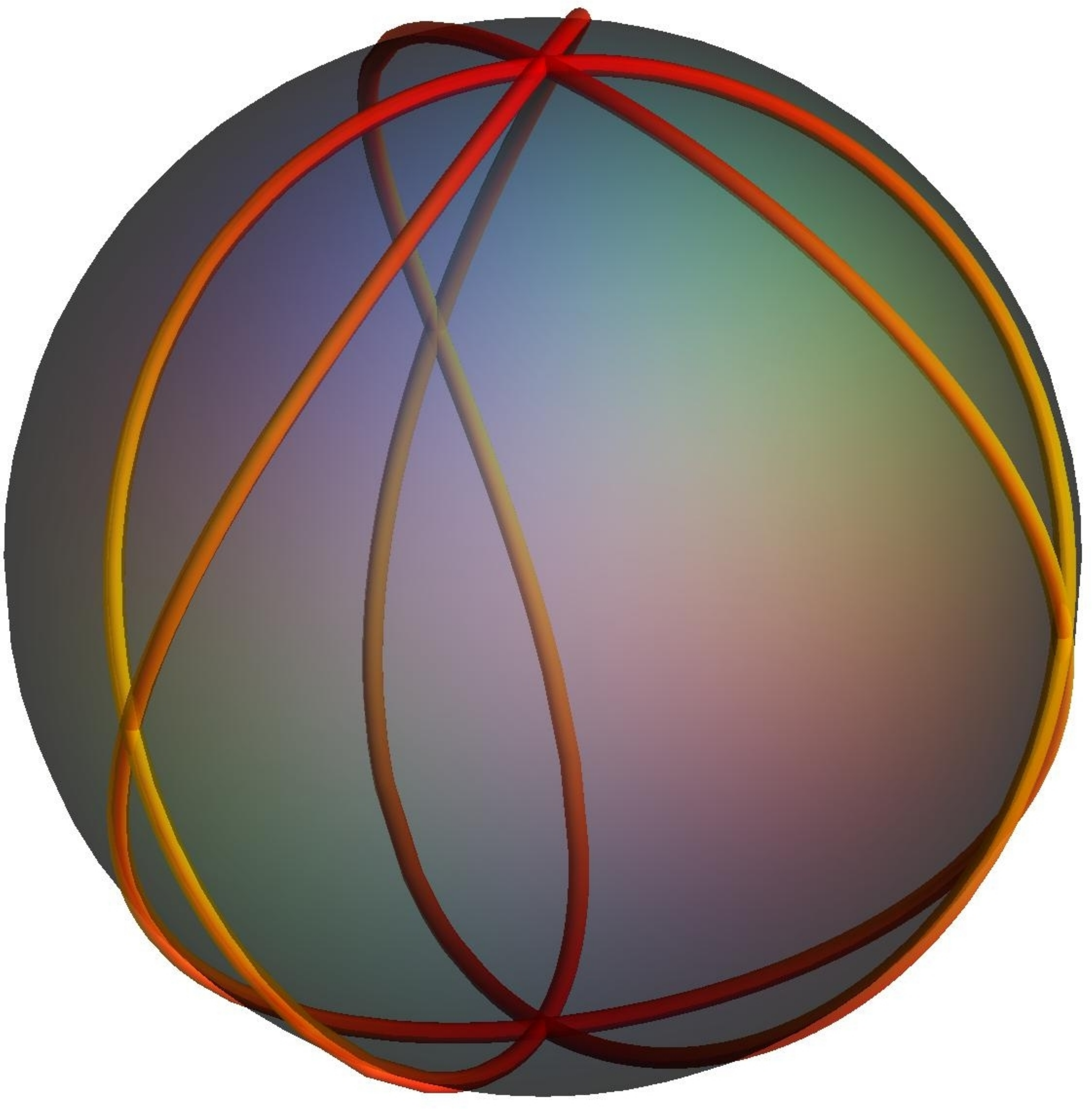}}
  \hfill
  \subfigure[$ R = 3.5$]{\includegraphics[scale=0.12]{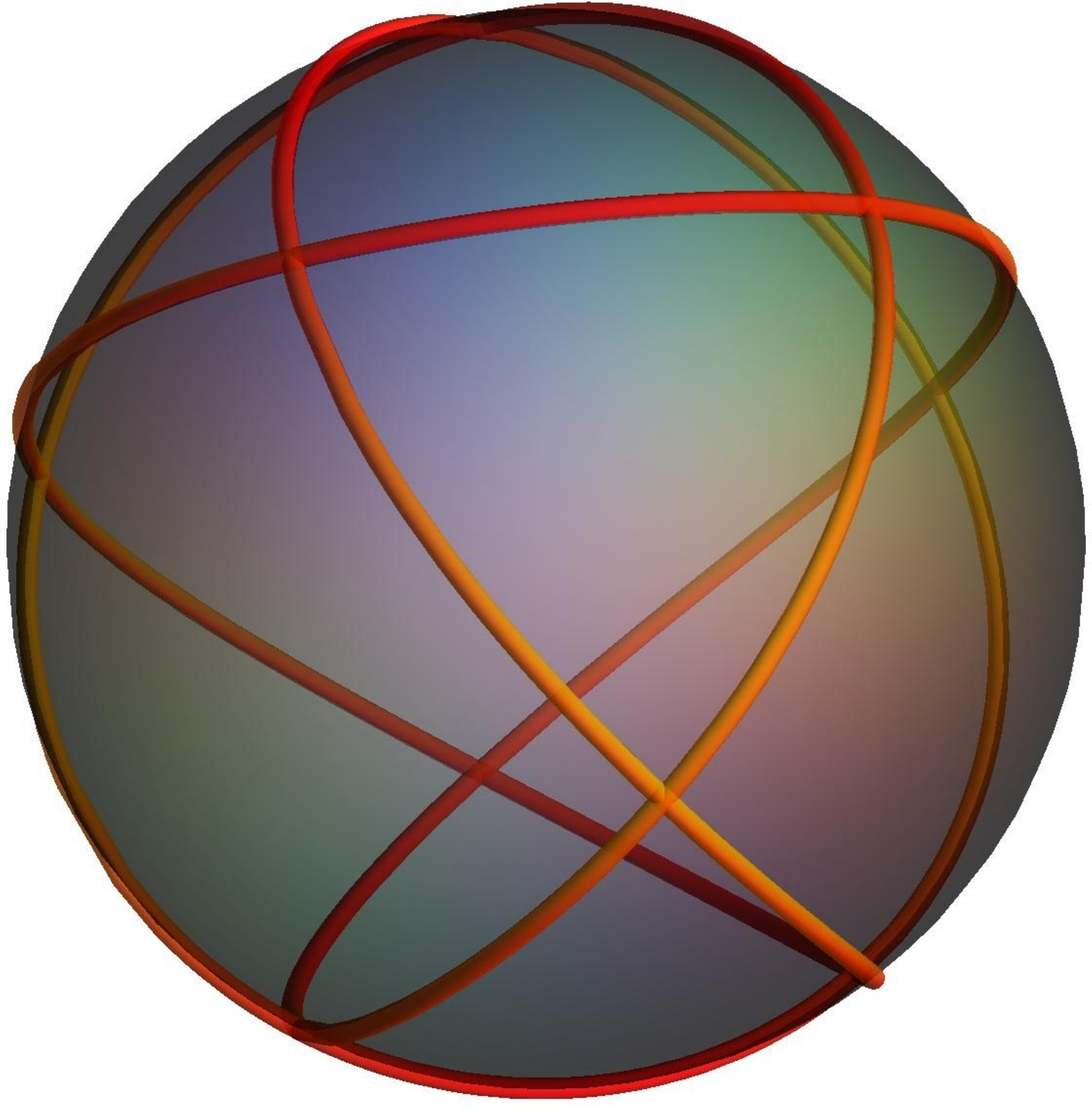}}
  \hfill
  \subfigure[$ R = 3.99$]{\includegraphics[scale=0.12]{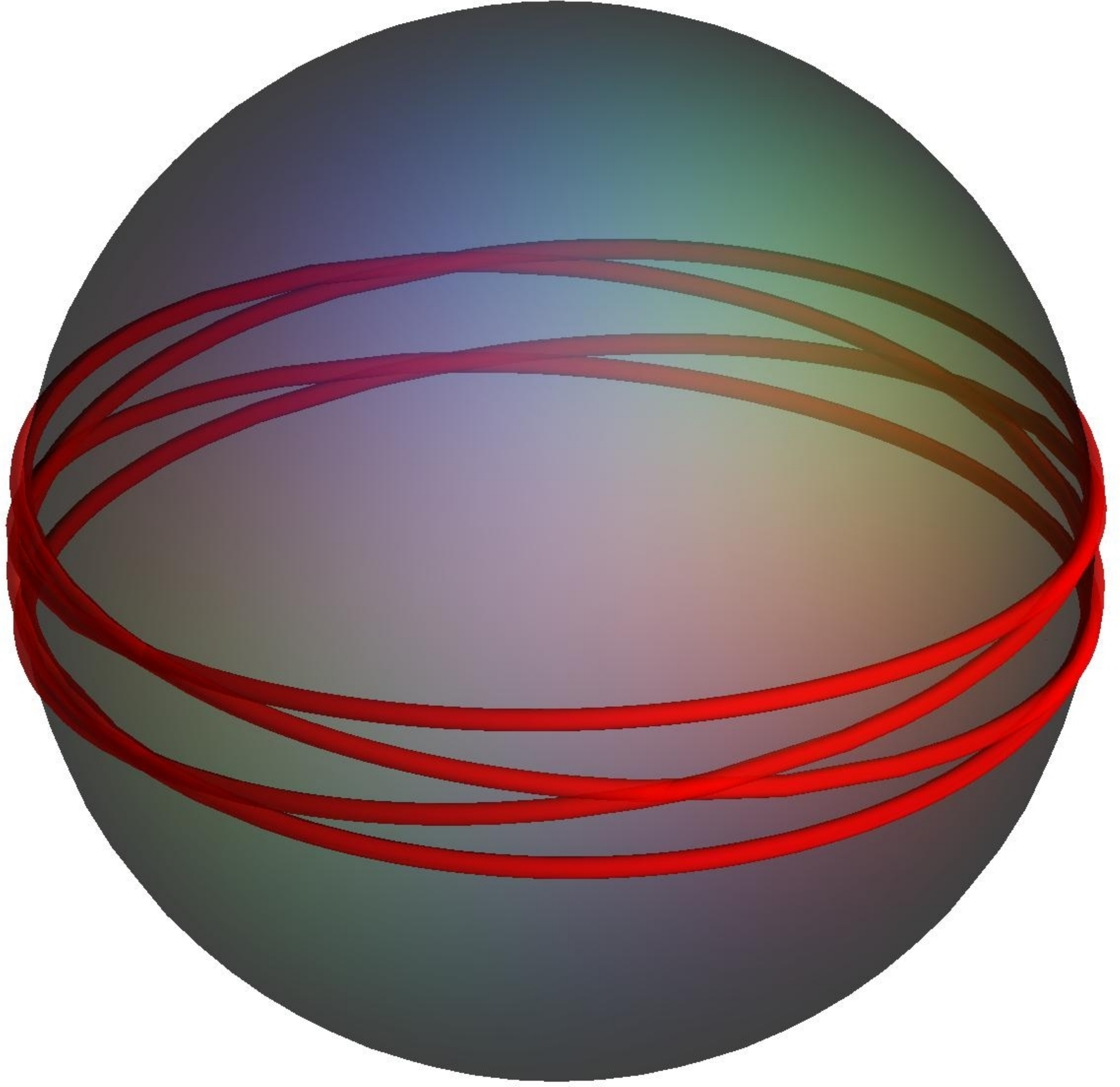}}\\
\includegraphics[scale=0.175]{fig1g.pdf}
\end{center}
\caption{\small (Color online) State with three-fold symmetry for $R$ in the
interval $[2,4]$: (a),(b),(c) increasingly deformed double covering of the
equator;  (d) pole crossing, three points at a time); (e) state making four
orbits of sphere; (f) quadruple covering of equator. As in Fig. 1 colors
represent the normalized local confining force.}\label{figure5}
\end{figure}

The $n$-fold symmetry states with $p=1$ and their descendants do
not exhaust all possible self-intersecting states.  If $R=2$, a
set of states  consisting of a double covering of a geodesic
circle ($p=2$) comes into existence with symmetry $n=3,5,7,\dots$.
The sequence $L_2$ with $n=3$ is illustrated in Fig. \ref{figure5}
for values of $R$ in the interval $[2,4]$.  As $R$ is increased,
the state with $n=3$ has least energy. As before, one can show
that it is the only stable member among these states. It also has
a lower energy than its counterpart with $n=2$, and $p=1$
illustrated in Fig. \ref{figure1}.

\vskip1pc \noindent The trajectory $L_2$ representing these states
in parameter space as $R$ is varied in the interval $[2,4]$, is
represented in Fig. \ref{figure2}(b) by a solid curve. When $R=4$ a
transition occurs to a state with a five-fold symmetry. The
sequence of trajectories $L_2, L_4,L_6, \dots\,$  is generated.
Their endpoints converge to the same point $P_{\infty}=(1,0)$ as
those of the sequence $L_1,L_3,L_5,\dots\,.$

\subsection{Loop Energy}

The bending energy Eq. (\ref{Htotsphere}) can be recast explicitly as a function
of $R$:
\begin{equation} \label{eq:Hstrong}
 H = \frac{16 \, n^2}{\pi R} \, {\cal K}[m] \,\left({\cal E}[m]+(m-1){\cal
K}[m]\right)+\pi R\,,
\end{equation}
where ${\cal E}[m]$ is the complete elliptic integral of the second kind
\cite{AbramStegun}, and $m$ is determined by solving Eq. (\ref{eq:varphiSm}).

\begin{figure}
\begin{center}
\includegraphics[scale=0.4]{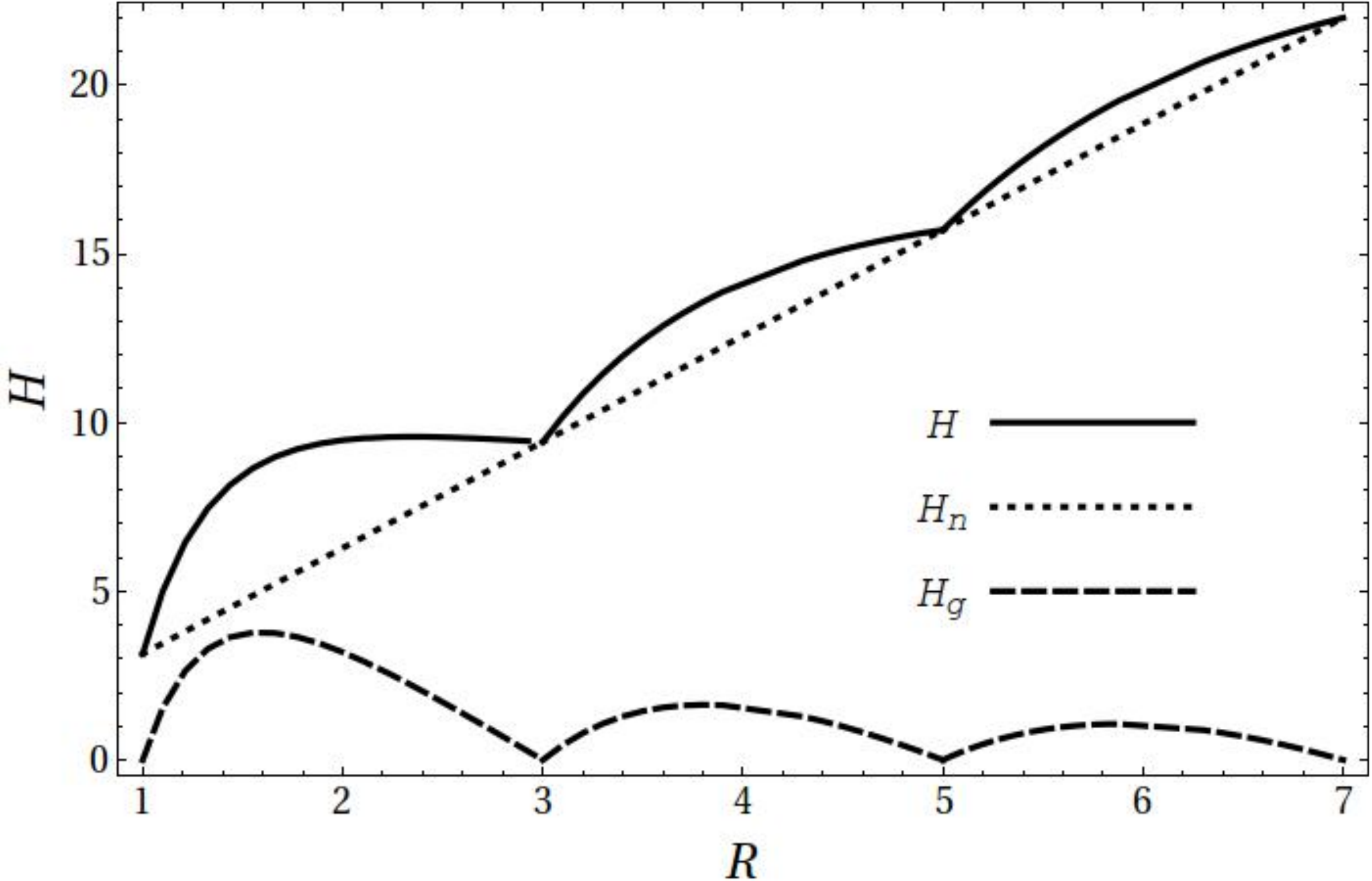}
\end{center}
\caption{\small The total energy of the $n=2$ ground state and its
descendants  as a function of $R$ is represented by a solid curve.
It is sum of geodesic and normal parts which are plotted separately
as a dashed curve and a dotted line. The normal energy grows
linearly and dominates in large loops. By contrast, the geodesic
energy is bounded, vanishing at odd integers where its derivative is
discontinuous. Its maxima between zeros decrease inversely with
$R$.}\label{figure6}
\end{figure}

\vskip1pc\noindent $H$ is plotted as a function of $R$ for the
ground state with $n=2$ and $p=1$ and its descendants in Fig.
\ref{figure6}.
Expanding expression (\ref{eq:Hstrong})
to second order in $\Delta R=R-1$ we reproduce
Eq.(\ref{eq:Hweak}).

\vskip1pc\noindent The normal energy, represented by the last term,
grows linearly with $R$ and is state independent. In contrast the
energy associated with geodesic curvature, represented by the
first two terms in Eq.(\ref{eq:Hstrong}), depends sensitively on
the state. It is bounded and its maxima fall monotonically as the
loop becomes large. It is simple to place a crude upper bound on
this falloff: one has $M \approx 2/R$, so that the $H_g \le 2/
R$.  Thus in a large loop, the geodesic contribution to the energy
is negligible compared to its state independent normal
counterpart.

\vskip1pc\noindent The geodesic energy vanishes when
$R=1,3,5,\dots$, where the loop collapses to a multiple covering
of a geodesic circle. The discontinuities in the derivative of $H$
with respect to $R$ at $R=3,5,\dots$, are directly associated with
the change of symmetry at these values of $R$. In the intervals
between consecutive values, $H$ increases linearly from zero and
rises to a maximum value before falling to zero, again linearly.\footnote { The initial linear behavior was described in
perturbation theory. The existence of a local maximum was not.}
Behavior is not symmetrical in these intervals. The slopes are
different at the two ends and the maximum is not centered.  While
the maxima are not exactly periodic, they do migrate to the center
of their respective intervals as $R$ increases. The existence of
these maxima  can be understood as a consequence of  the
incommensurability of the loop with geodesic behavior. The
strongly asymmetrical initial maximum is associated with the
development of overhangs on the loop before it crosses the poles.
As the loop becomes longer this incommensurability plays a
diminishing role.

\vskip1pc\noindent This qualitative behavior is repeated for the
odd ground states $n=3$, $p=2$ and its descendants, as well as the
excited counterparts of these states. Strikingly, the energy of any
closed equilibrium loop tends to a common value in the limit,
independent of the state, that completely dominated by the normal
energy. The finite gap between the ground state and excited states
state disappears.

\vskip1pc \noindent If $R-1$ is not small, the stability analysis
is rather more complicated than the one presented in the context
of weak confinement.  The results  of Ref. \cite{Stability}
in another context implies that, although the details differ,  the
conclusions for weak confinement continue to hold for values of
$R$ before the onset of self-intersection and only the state
$n=2,3,4,\dots$, $p=1$ are excited. A detailed analysis of
stability has yet to be performed beyond this point. One would,
however, expect that the instabilities persist.

\begin{figure}
\begin{center}
\includegraphics[scale=0.4]{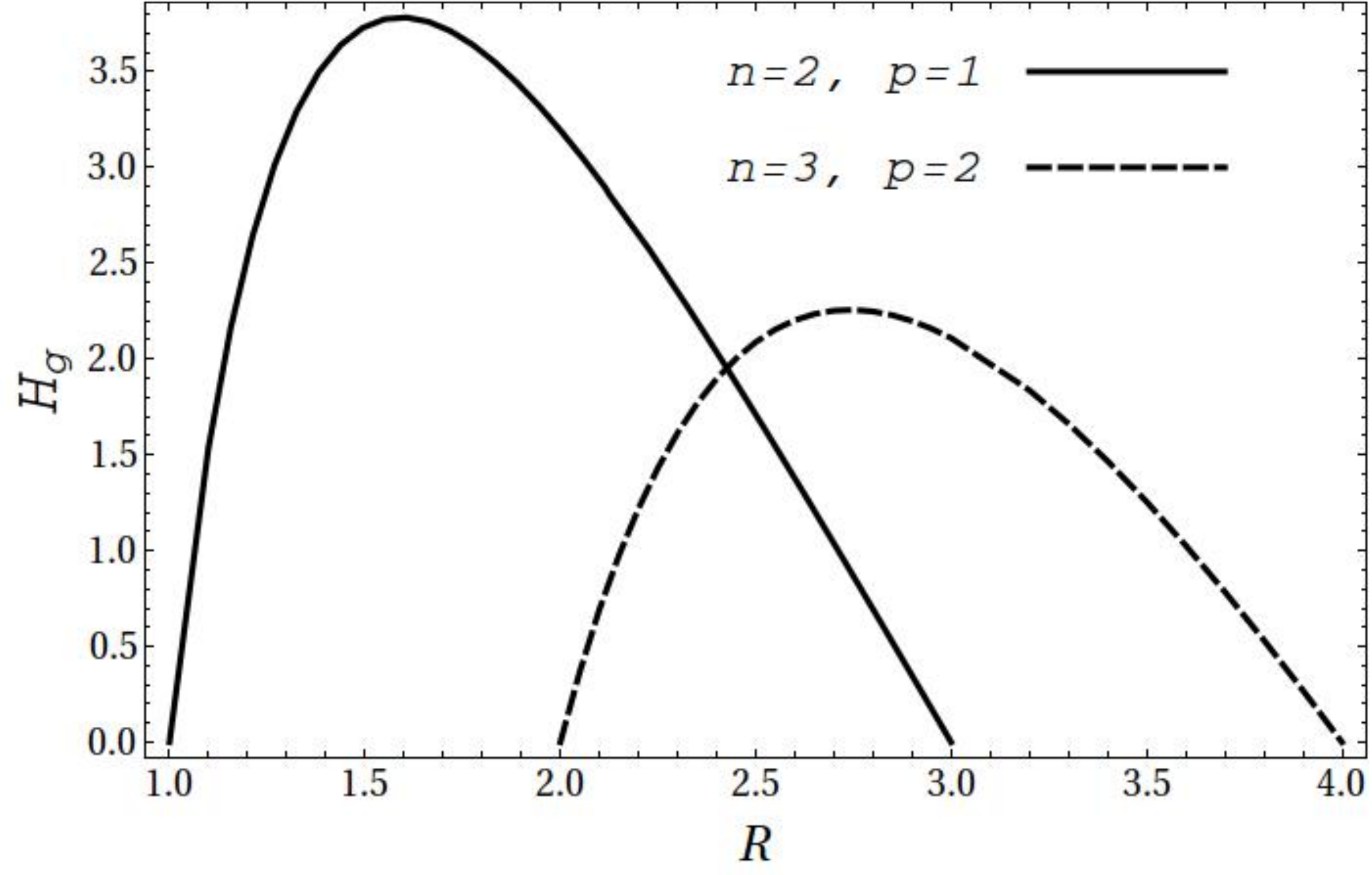}
\end{center}
\caption{\small Comparing the geodesic energies of the states with
two  and three fold symmetries in an interval of $R$ in the
neighborhood of $R=2$ (solid and dashed lines respectively). The
geodesic energy, and thus the total energy, of the three-fold is
lower than that of the two-fold in a finite interval of $R$
beginning at $R=2$.}\label{figure7}
\end{figure}

\vskip1pc \noindent
The stability of the $n=2$, $p=1$ ground state and its descendants needs to be
reassessed when $R\ge 2$. For now one has to accommodate the existence
of a set of states described by $n=3,4,5,\dots\,$ $p=2$ and their
descendants. In particular, within a finite interval of values of $R$ above $R=2$,
the state $n=3$, $p=2$, represented in  Fig. \ref{figure5}  has lower energy
than the state
with $n=2$ and $p=1$ represented in Figure \ref{figure1}).
See Fig. \ref{figure7}.  The difference in energy is associated
with the significant geodesic curvature of the latter when $R\approx 2$. Its
descendants with $n=5,7,\dots$ also will have lower energy than their even
counterparts when $R\approx 4,6,\dots$.
%In each of these regimes one needs to reassess the stability of the latter.

\begin{figure}
\begin{center}
  \subfigure[$t = 0$]{\includegraphics[scale=0.12]{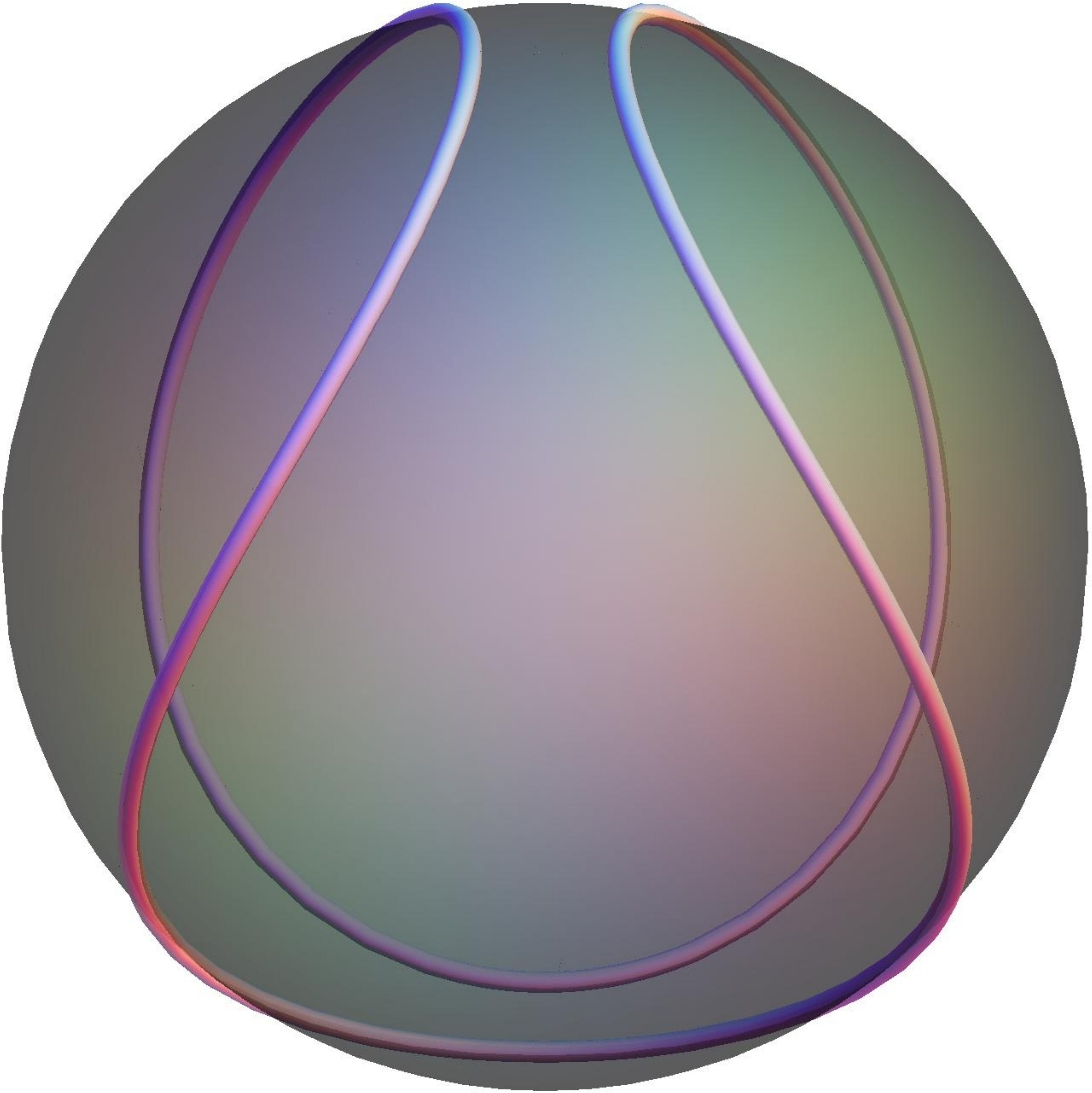}}
  \hfill
  \subfigure[$ t = 0.2$]{\includegraphics[scale=0.12]{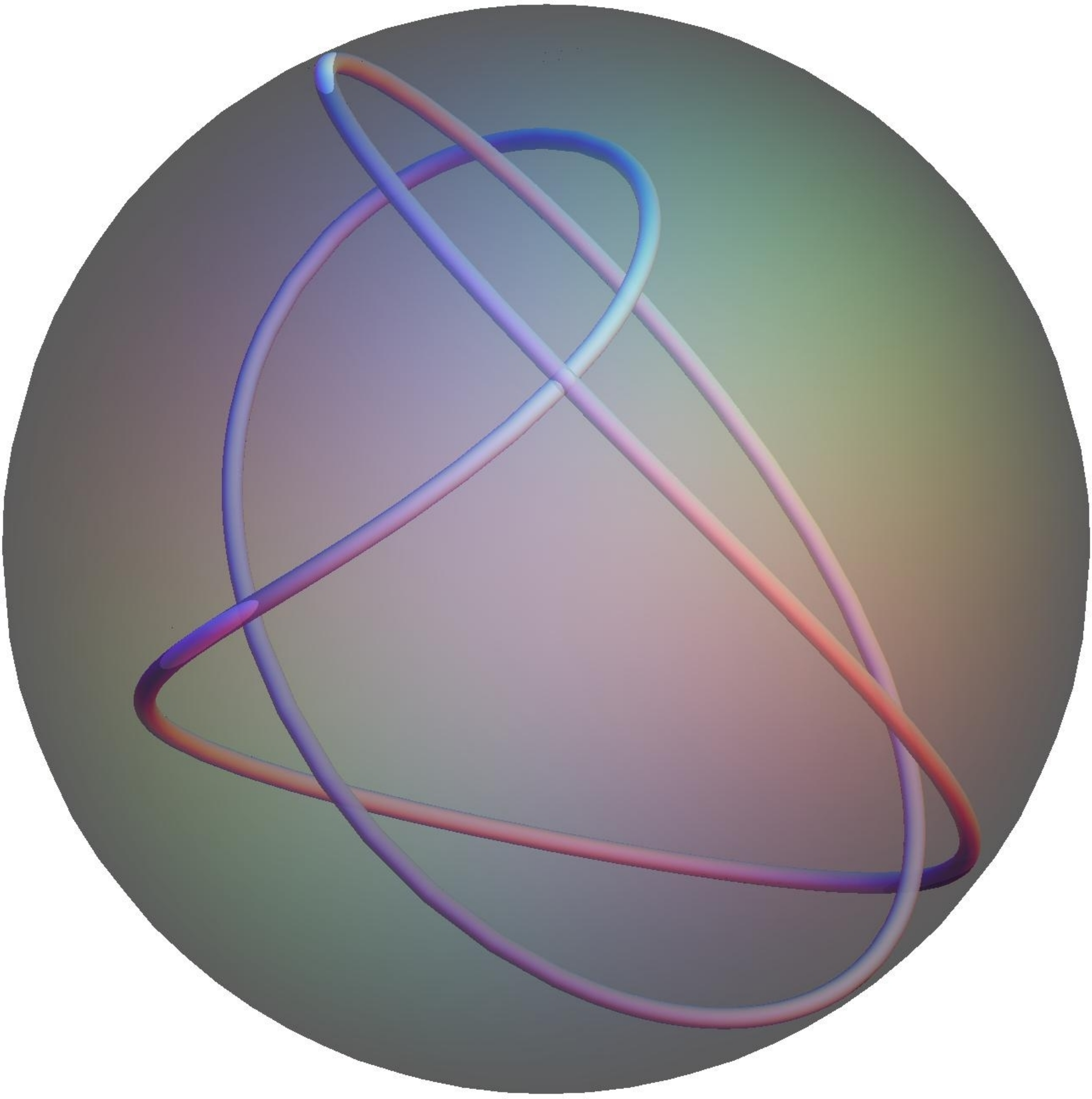}}
  \hfill
  \subfigure[$ t = 0.4$]{\includegraphics[scale=0.12]{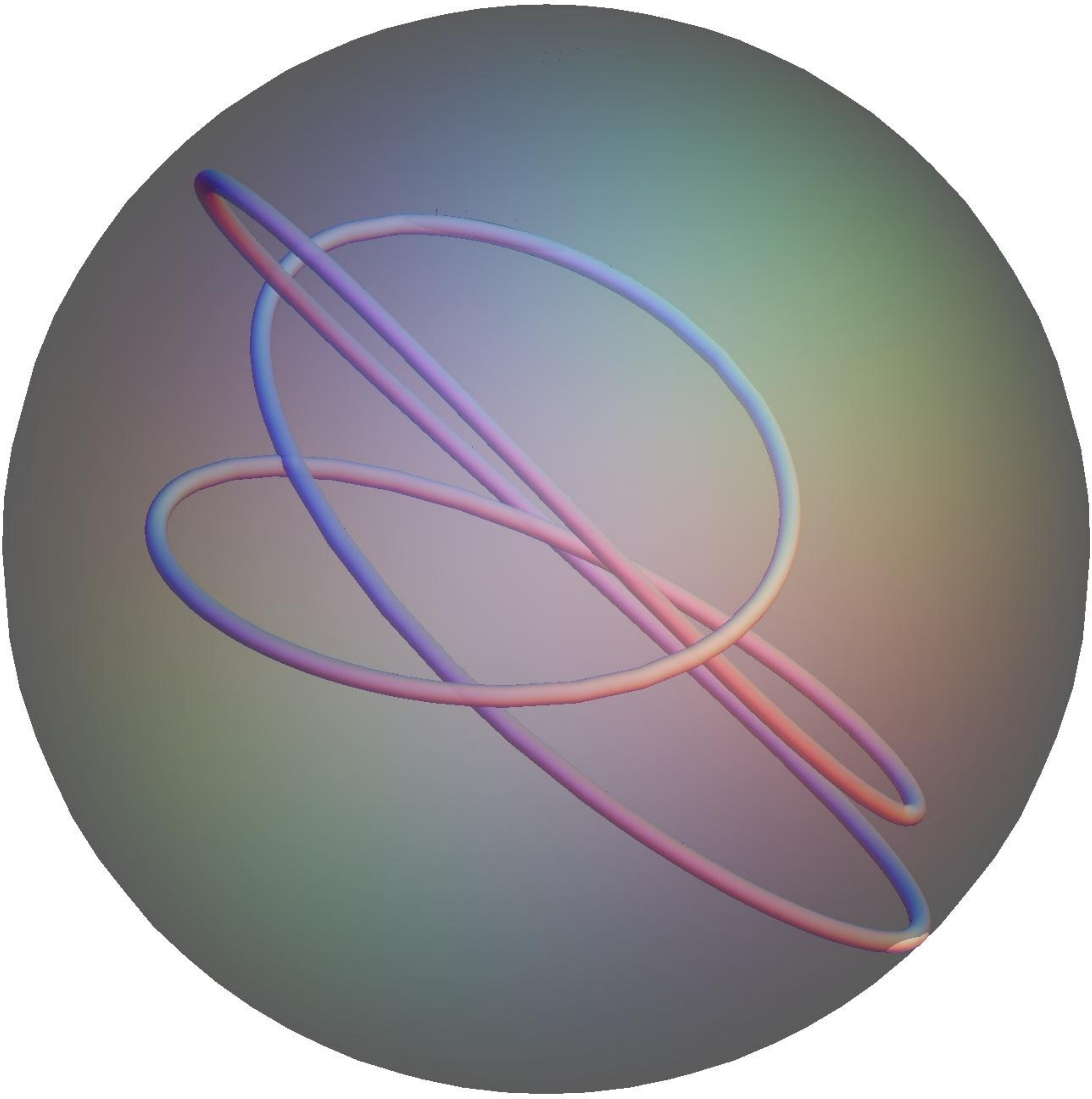}}\\
  \subfigure[$ t = 0.6$]{\includegraphics[scale=0.12]{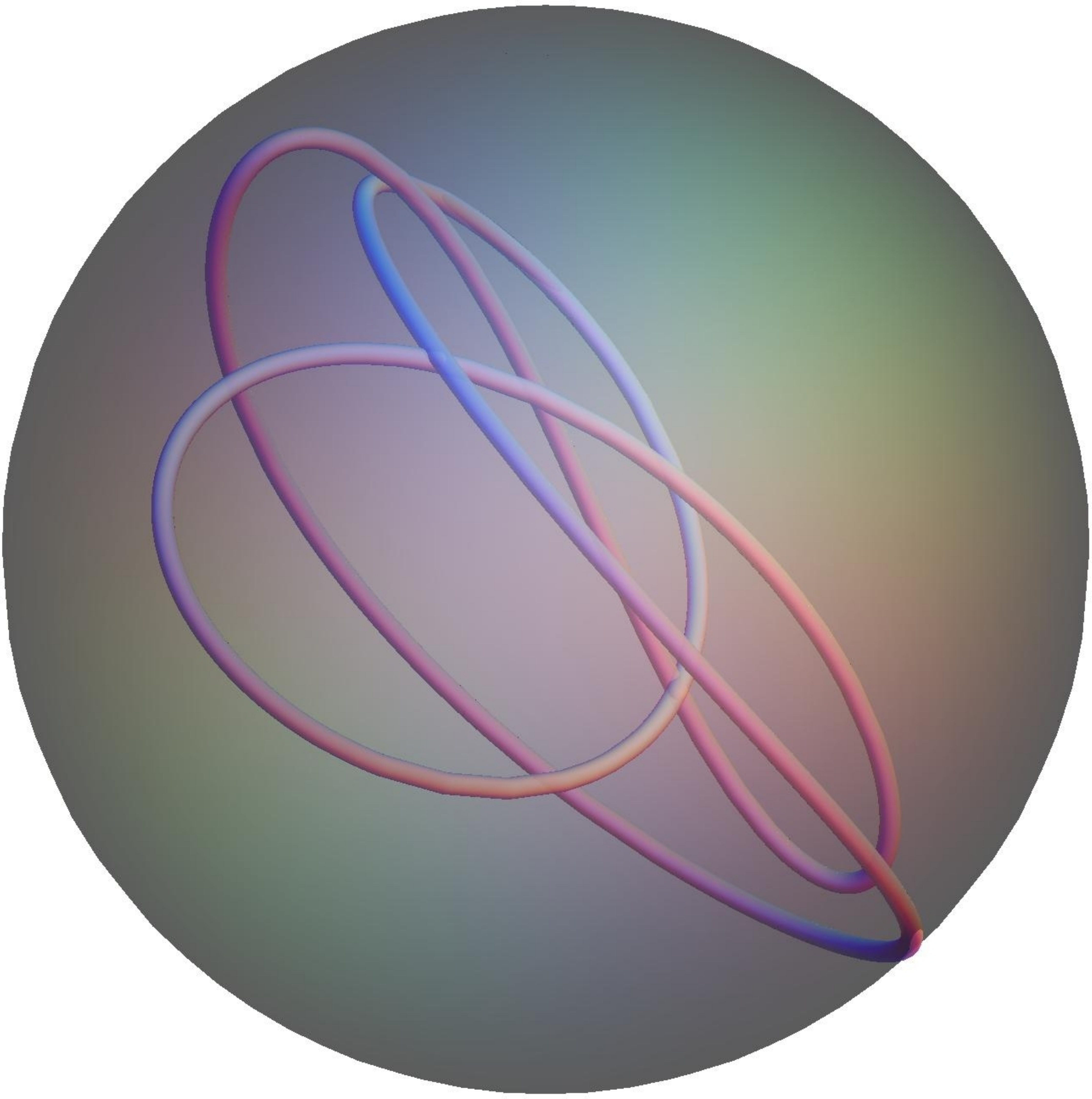}}
  \hfill
  \subfigure[$ t = 0.8$]{\includegraphics[scale=0.12]{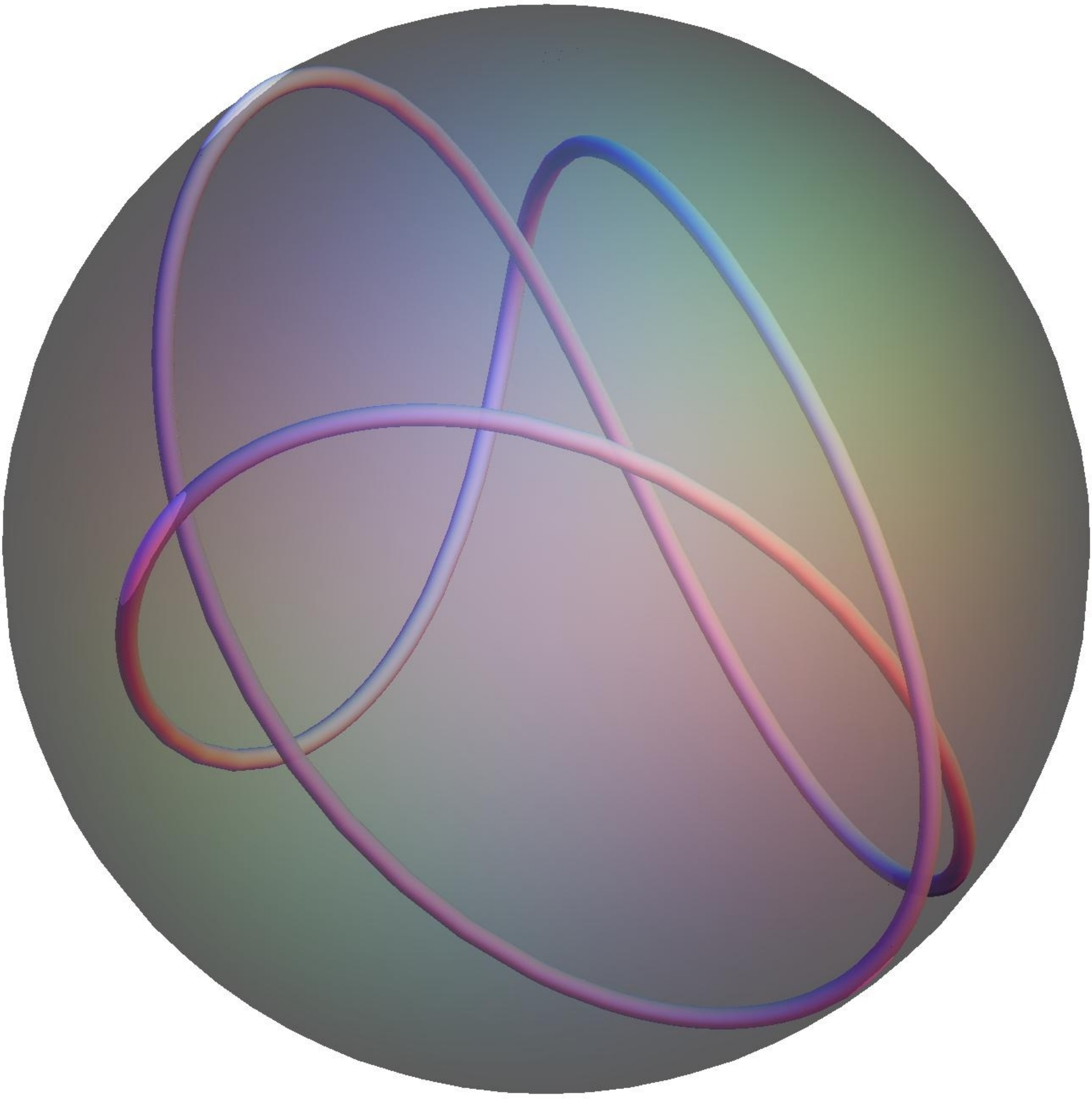}}
  \hfill
  \subfigure[$ t = 1$]{\includegraphics[scale=0.12]{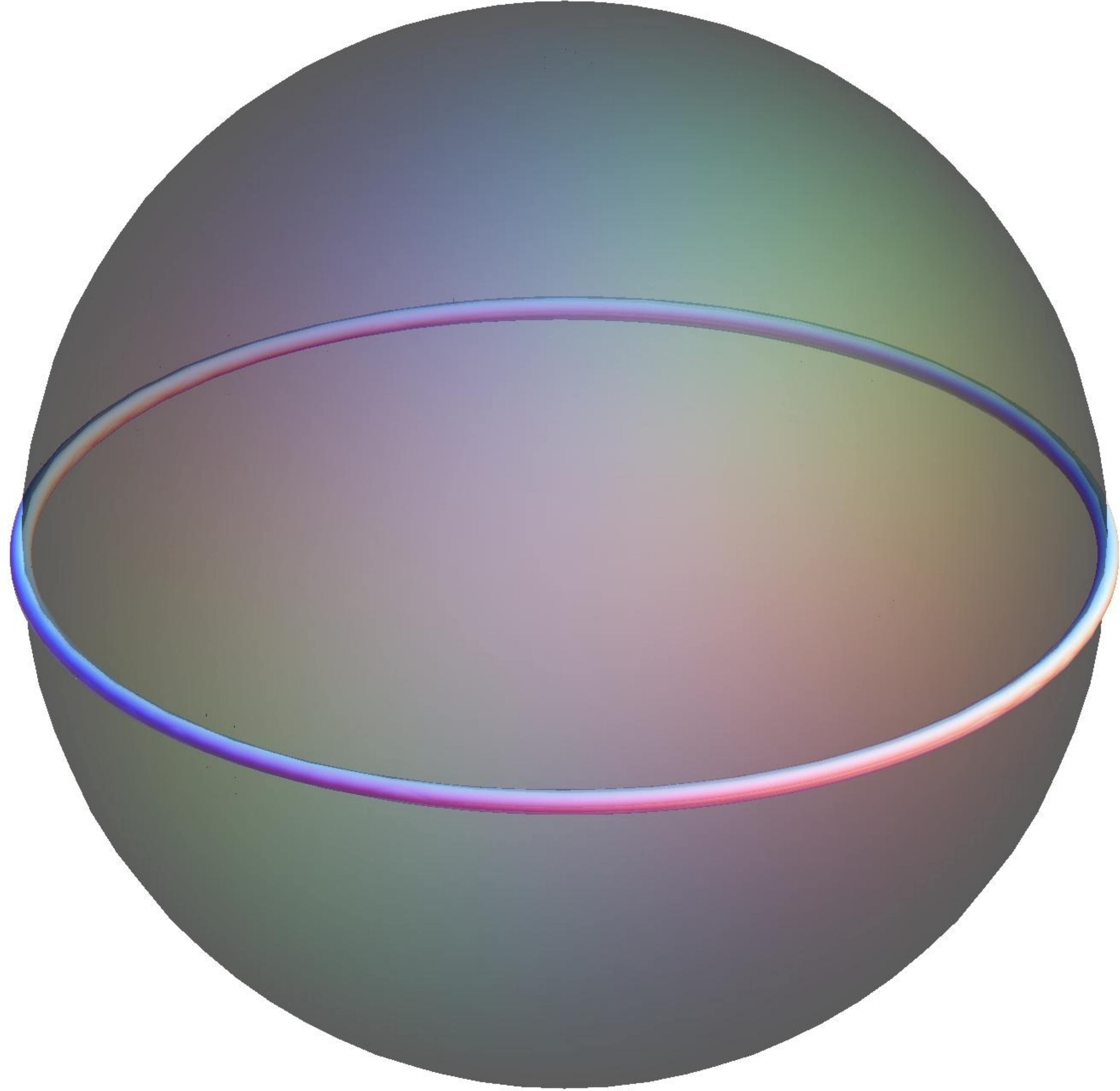}}
\end{center}
\caption{\small (Color online) Homotopy connecting the two states $n=2, p=1$ and
$n=3, p=2$
with $R=2$.}\label{figure8}
\end{figure}

\begin{figure}
\begin{center}
\includegraphics[scale=0.4]{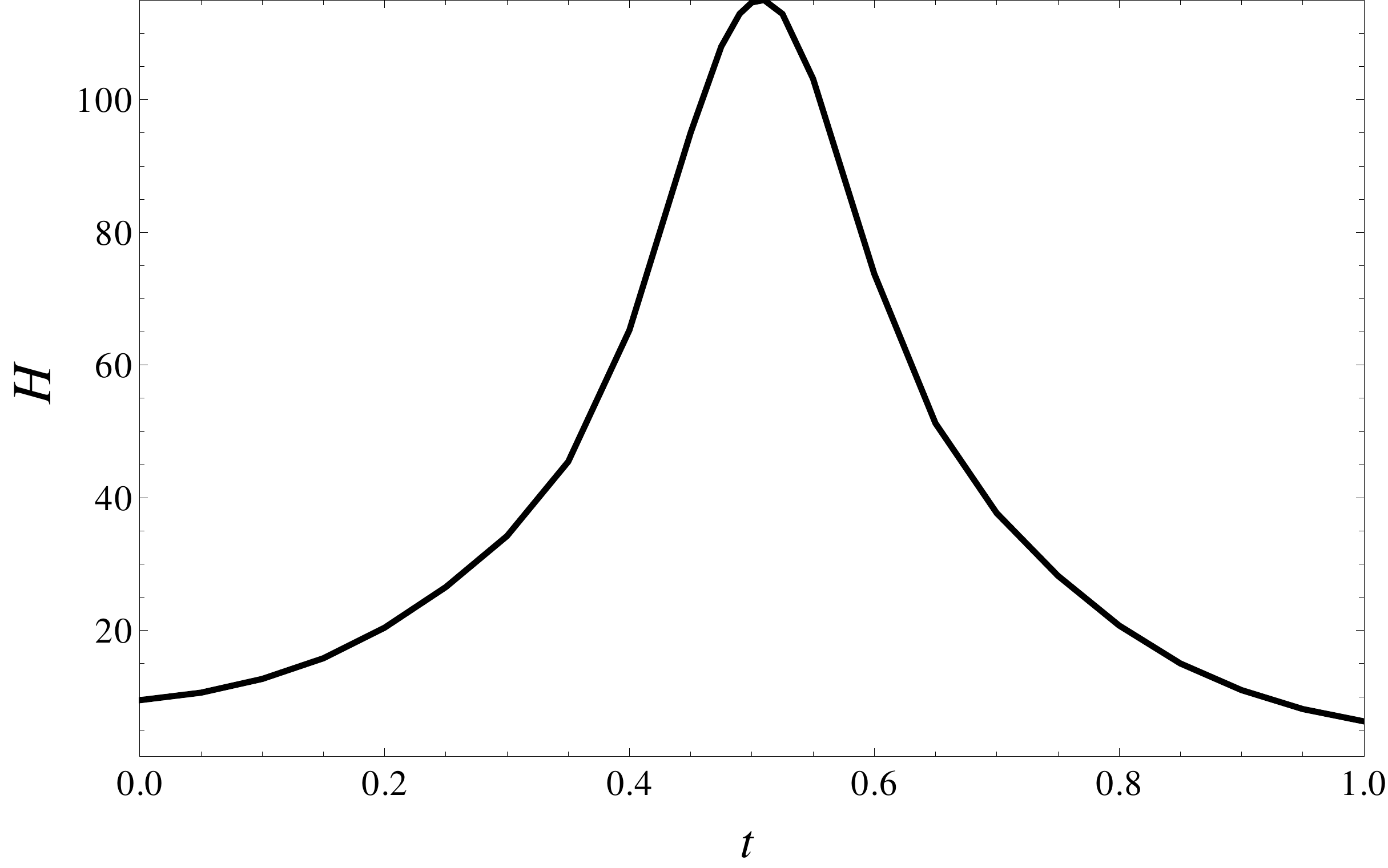}
\end{center}
\caption{\small Total energy along a homotopy interpolating between two stable
ground states. An energy barrier always exists between the two.}\label{figure9}
\end{figure}

\vskip1pc \noindent There are, of course, no smooth deformations
taking one from a loop with $p=1$ to one with $p=2$ that remain
attached to the sphere. A hairpin costing an infinite bending energy
will necessarily always form.  However, it is possible to sidestep
this topological obstruction by permitting the loop to detach into
the interior.  While a rigorous stability analysis is beyond the
scope of this paper,  we will argue that a finite energy barrier
will always separate the two surface-bound states. This involves the
construction of a natural homotopy connecting these states.
Let
\begin{equation}
 {\bf Y}_t = (1-t) {\bf Y}_0 + (1-t) {\sf R}(\omega) {\bf Y}_1\,,
\end{equation}
where we represent by ${\bf Y}_0$ the initial state with $n = 2,
p=1$ and by ${\bf Y}_1$ the final state with $n = 3, p=2$. These
interpolating non-equilibrium states will also be confined within
the sphere due to the convexity of the latter.  The two boundary
states have $R = 2$.\footnote {As written down, the homotopy does
not preserve length.  This can be achieved by stretching ${\bf
Y}_0$ and ${\bf Y}_1$ appropriately at intermediate values of $t$.}
A rotation ${\sf R}(\omega)$ of the final state about the axis of
symmetry is introduced in order to initialize the pointwise
sum so as to avoid the development of cusps.
Intermediates in this homotopic sequence are illustrated
in Fig. \ref{figure8} with the constant choice $\omega= \pi / 6$.
The bending energy is plotted in Fig. \ref{figure9} as a function
of $t$. It exhibits a finite potential barrier separating the two
states. While this does not prove that the $n=2$ state is stable
in this regime, it does suggest that it is.

\vskip1pc\noindent One cannot rule out the existence of
semiattached equilibrium states of the loop. We were, however,
unable to construct any such states. If they do exist, one would expect
them to be unstable.

\subsection{Transmitted forces}

Using Eq. (\ref{Eq:kappagab}),  the local force per unit length
transmitted to the sphere is given  explicitly as the following function of $s$,
\begin{equation}
\label{eq:Fstrong}
\lambda (s) = \frac{1}{2} \kappa^2_g + \sigma = q^2 \left(2 \,
\mbox{dn}^2\left(q\, s, m\right)-1\right)\,,
\end{equation}
where $\text{dn}^2[u,m] = 1-m \,\text{sn}^2 \,[u,m]$
\cite{AbramStegun}. Whereas the transmitted force depends on the
local value of $\kappa_g$,  both $\kappa_g$ and $\sigma$ depend on
the boundary conditions associated with closure. Thus $\lambda$
does depend on the global loop geometry. Note that the expansion of the
expression in (\ref{eq:Fstrong}) to second order in $\Delta R=R-1$ reproduces
the result for  weak confinement (Eq. (\ref{eq:Fweak})).

\begin{figure}
\begin{center}
  \includegraphics[scale=0.4]{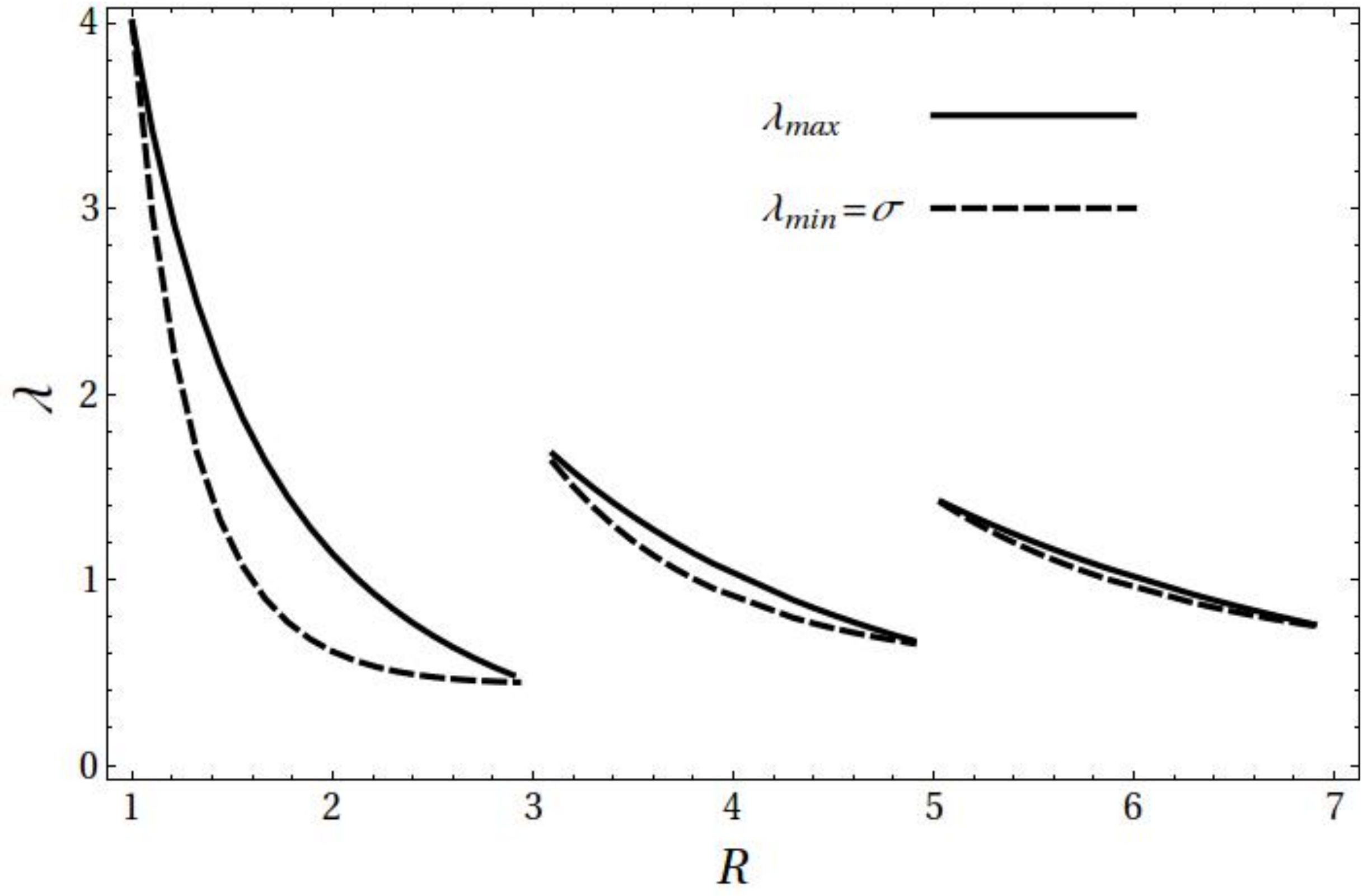}
\end{center}
\caption{\small Transmitted force. Maximum (solid line) and minimum (dashed
line) values of $\lambda$ as a function of $\Delta R$ in the two-fold ground
state with $n=2$, $p=1$, $\lambda$ in the two-fold ground state and its minimum
energy descendants.
}\label{figure10}
\end{figure}

\vskip1pc \noindent
In Fig. \ref{figure10} the maximum and minimum values of $\lambda$ have been
plotted as a function of $R$ for the two-fold ``ground'' state and its
descendants. It is positive everywhere  with the  minimum given by $\sigma$. One
observes that $\lambda$ does not behave in a monotonic way, suffering a positive
discontinuity whenever $R=p$, where $p$ is an odd integer. This is associated
with the transition from orbital to oscillatory behavior at these values and is
the analog of the Euler instability described in the discussion of a weakly
confined loop. If the unstretchability constraint is relaxed the discontinuity
will be smoothed but the jump will persist.  Within the individual intervals
between discontinuities both the maximum and minimum values decrease
monotonically with $R$; this is associated with the reduction in the force once
buckling into a state of oscillation has occurred.  Curiously, whereas the
maxima within these intervals decrease with $R$, the corresponding minima
increase. This occurs in such a way that the mean of their values increases
monotonically. Both maxima and minima approach the value $\sigma = 1$
asymptotically.

\begin{figure}[htb]
\begin{center}
\subfigure[Oscillatory State $n = 5$, $p =
1$]{\includegraphics[scale=0.15]{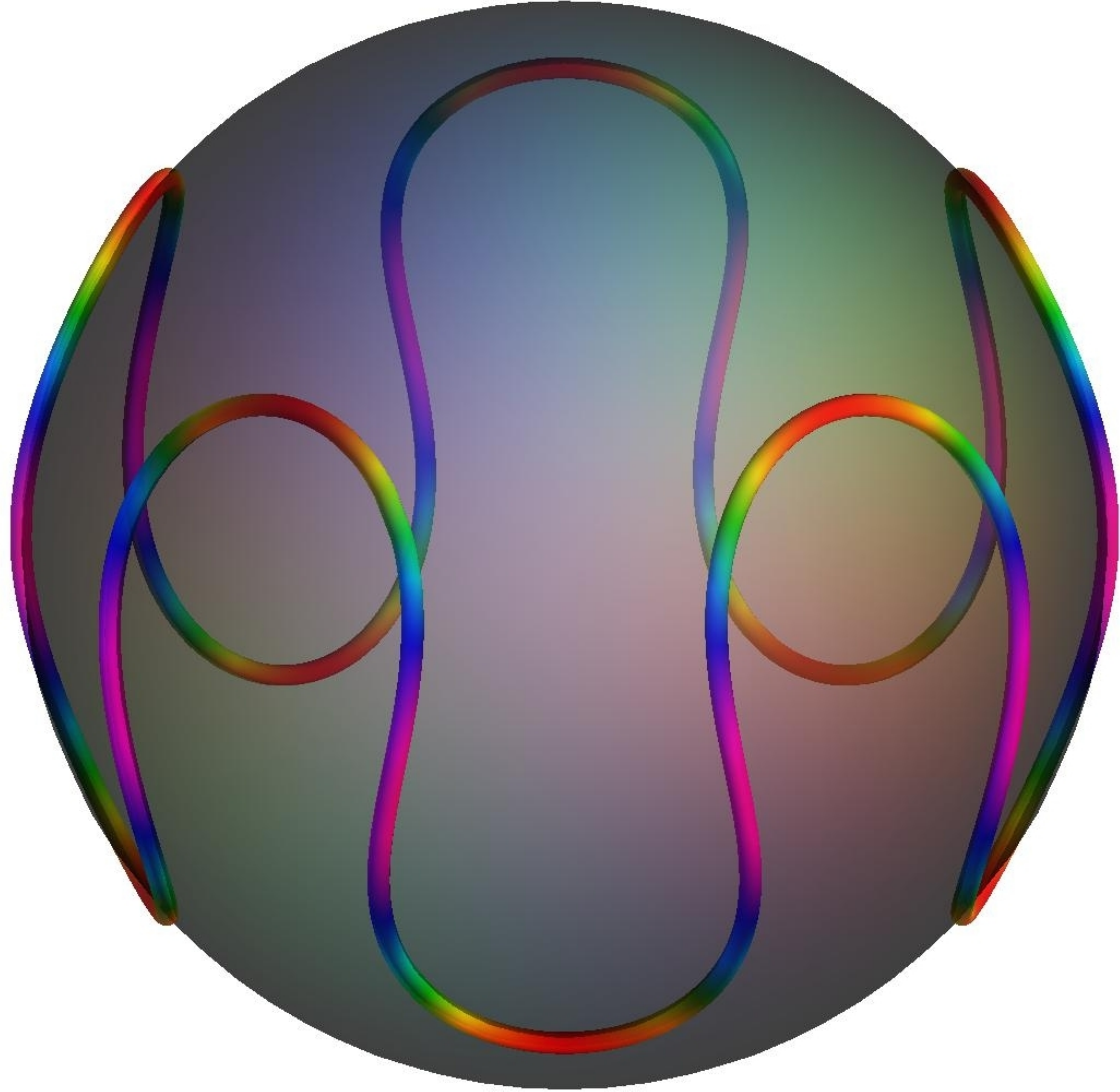}}\qquad \qquad \qquad \qquad \qquad
\subfigure[Orbital State $n=2$,
$p=1$]{\includegraphics[scale=0.15]{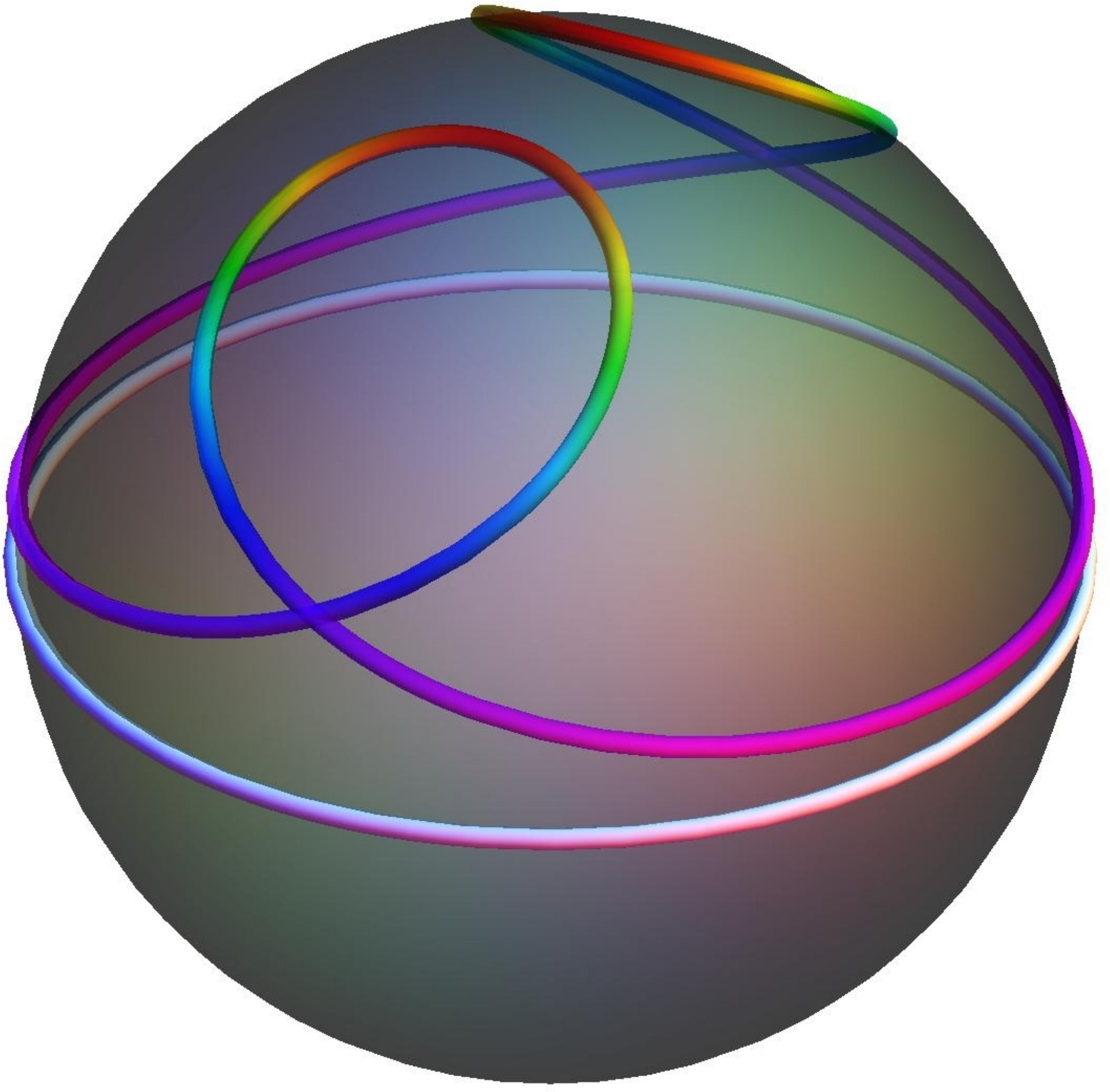}}\\
\includegraphics[scale=0.15]{fig1g.pdf}
\qquad \qquad
\includegraphics[scale=0.15]{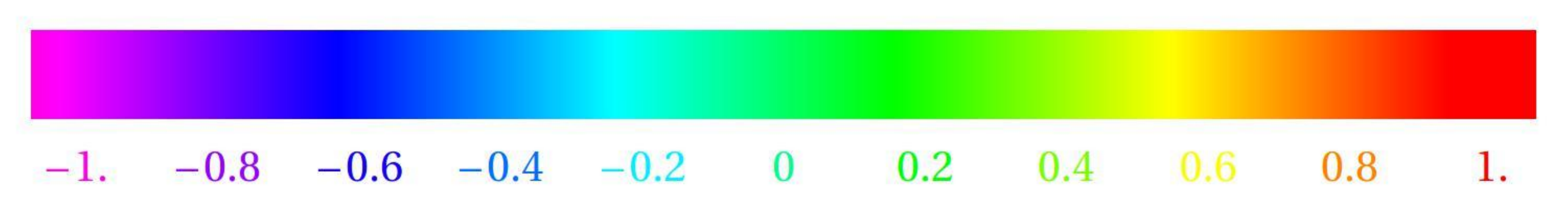}
\end{center}
\caption{\small (Color online) (a) The state $n=5$, $p=1$ with $R = 2.536$.
Above this value of $R$, $\lambda$ may be negative in places. (b) The
state $n=2$, $p=1$ with $R = 1.85$ inhabits a single hemisphere. These orbital states always
have intervals where $\lambda$ is negative.} \label{figure11}
\end{figure}

\vskip1pc \noindent Recall that the transmitted force is bounded from below by
$\sigma$ (see Eq.(\ref{lambdasphere})). Thus, if $\sigma > 0$, then $\lambda$ is
also; the confined loop will then push on the sphere everywhere. If,
however, $\sigma < 0$ then $\lambda$ may change sign along the
curve. While $\sigma$ is positive in the ground state and its
descendants, it may become negative in the excited states of
sufficiently long loops. In particular, we find that all states
with $n\ge 5$ and $p=1$ exhibit regions in which $\lambda$ turns
negative at values of $R$ below the onset of self-contact. When
$n=5$ and $p=1$, this occurs when $R \geq 2.536$ (see fig
\ref{figure11}(a)).  Such states are, however, unstable.
It may appear counterintuitive that $\lambda$ can be negative. It
has to be remembered, however,  that while $\lambda$  depends only
on the local geometry, this geometry  is determined by the global
behavior of the loop through the boundary conditions. If one cuts
a loop somewhere, the state will immediately relax into a geodesic
state with  constant positive $\lambda$ everywhere. The existence
of regions along the loop where $\lambda<0$ does not necessarily
signal a tendency to detach.  This will depend on details of the
energy landscape.

\vskip1pc \noindent
The states examined here are those consistent with the bound Eq.
(\ref{eq:parabola}).  As was pointed out, states
that violate this bound are always unstable with respect to the
unbinding of the loop into the interior. Such a state is illustrated
in Fig. \ref{figure11}(b), making two self-intersecting orbits in the
northern hemisphere. The curvature is high within the two orbits so that this state is
evidently an excited state of the loop. The transmitted force  $\lambda$ is negative
everywhere below the circle with $\kappa_g =\sqrt{-2 \sigma}$ (see
Fig. \ref{figure11}(b)). This suggests that the likely mode of instability will
involve the loop unbinding along regions where $\lambda$ is negative, allowing
the loop to relax toward the ground state $n=2$, $p=1$ through the unwinding of
the high curvature orbits.

\begin{figure}[htb]
\begin{center}
\includegraphics[scale=0.4]{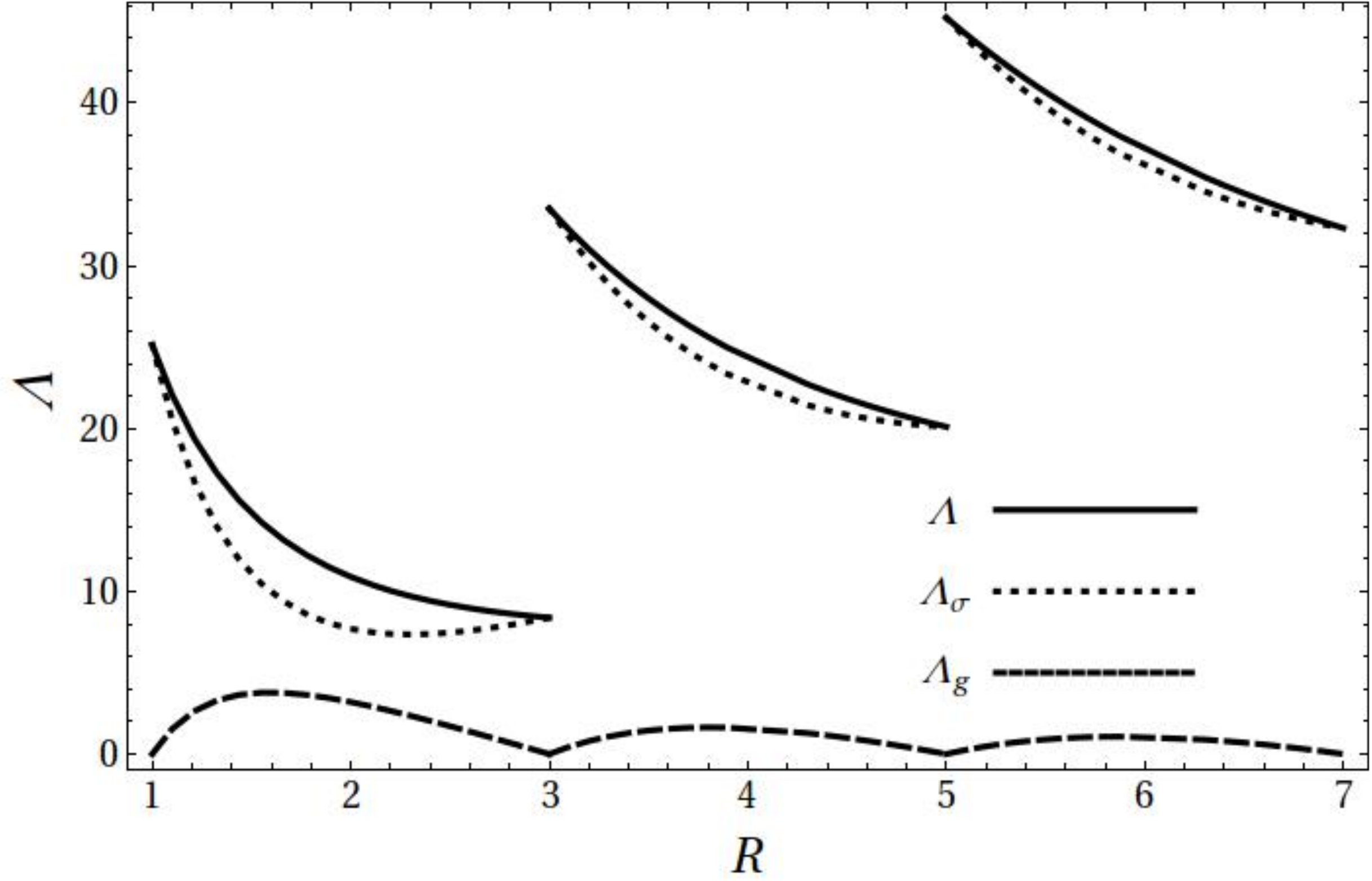}
\end{center}
\caption{\small $\Lambda$ vs. $R$ (solid line).  The dashed line indicates the
contribution from geodesic curvature $\Lambda_{\textit g}$. The dotted line is the contribution
$\Lambda_\sigma = 2 \pi R \sigma (R)$.}
\label{figure12}
\end{figure}

\vskip1pc \noindent Intuitively, one might have expected the total
force transmitted to the sphere, $\Lambda = \int ds \lambda$ to
increase with loop size. This is not generally the case, except in
the limit of large loops, where $\Lambda$ increases linearly with
$R$, a consequence of the asymptotic behavior of $\sigma$. Notice also
that $\Lambda$ is not the same as the energy $H$ (divided by the
radius of the sphere) for finite values of $R$. One has $\Lambda = H
- 2 \pi R (1-c (R))$. In the limit, however, the two do coincide.
$\Lambda$ is plotted as a function of $R$ in Fig. \ref{figure12}.

\section{Discussion} \label{Discussion}

The confinement of a closed semiflexible loop by a surface has
been examined. The constraint on the spatial degrees of
freedom of the loop is enforced in the variational principle using
the method of Lagrange multipliers. The loss of  Euclidean invariance of the energy of the
unconstrained loop under confinement is quantified by the Lagrange multiplier which,
in equilibrium, gets identified as the force transmitted to the surface
by the confined loop.

\vskip1pc \noindent We have focused on a description of the ground
state of a confined loop. If the loop is short this consists of an
oscillation about a geodesic circle, exhibiting a two-fold
dihedral symmetry.  This is the only stable state of the loop in
this regime. The description of the ground state and its
excitations gets more complicated when the loop length is
increased. When $R$ is increased above $2R_0$,  a new set of
states comes into existence that oscillates about a double
covering of a geodesic circle. Among these the state with
three-fold symmetry has lowest energy; within a finite band of
values of $R$, this state also replaces the two-fold as the ground
state. The two states remain separated by an energy barrier so
that the two-fold remains stable. As the loop size increases these
states get replaced by descendants with higher $n$-fold symmetries
which alternate as ground states. Both the energy and the transmitted
force suffer discontinuities associated
with changes of symmetry. They do not generally
increase monotonically with loop size except asymptotically. In this limit the
energy gap between the ground state and excited states disappears.

\vskip1pc \noindent Our examination of the confinement of a loop
within a spherical cavity might lead one to expect that the
equilibrium states of a confined loop will always attach. However, other
confining geometries display very different behavior. Consider,
for example, the confinement of a loop by a cylinder of smaller
radius. Analogs of the $n$-folds exist but, in general, they do
not provide the ground state
% corresponding ground state of the loop, however,
%is a completely bound elliptical deformation of the circular loop, a state
%without any spherical analogue. The corresponding two-fold has higher energy
%and is, at least in the regime of small loop size,  unstable with respect to
%deformations.
as one can easily verify by playing with metal rings in a wastepaper
basket.
%the elliptical shape undergoes a transition to a deformed circle aligned along the %axis of the
%cylinder when the length of the loop is increased.
%The ground state will not generally be a fully attached state.
%The contacting two-fold appears as a stable excited state.
The details are surprisingly complicated. In this context, it would
be interesting to understand how the spherical equilibrium states
``evolve'' under elliptical deformations of the sphere, or what
effect surface irregularities or pores have on the confinement
process.

\vskip1pc \noindent
When contact with the confining surface is incomplete, one needs to address the
issue of boundary conditions at points of contact. If the contact is  due
entirely to geometric hindrance, the boundary conditions at isolated points of
contact  are simple: the tangent vector to the curve is tangent to the surface
at these points. If contact extends over a finite region, however, in addition to the tangent vectors the
curvatures will also be continuous at the boundary of the region of
contact.  There may also be
a tendency to either promote or inhibit adhesion to the confining surface. A
simple way to accommodate such interactions is to introduce a local contact
energy, proportional to the length of the contact interval. The boundary
conditions will reflect this additional energy.  While the curvature will suffer
a discontinuity analogous to the discontinuity associated with the contact line
bounding the region of contact between a fluid membrane  and an attractive
substrate \cite{Seifert,SeifertLipowsky,Adh,Contour}, this is not the full
story. The normals may rotate about the tangent vector, aligning themselves
with some preferred direction on the confining geometry. The extension of our
framework to accommodate contact energies will be presented
elsewhere.

\vskip1pc \noindent In this paper the role played by the confining surface has been passive: we have not considered the possibility that it may deform in reaction to the presence of the confined polymer. Surface deformations due to membrane bound polymers may also play a role in shaping the morphologies of biological membranes.  The forces transmitted to the membrane will now provide a source for the surface stress. Understanding this coupling
poses technical challenges. What is clear is that there will
be interesting physics associated with the competing
elastic energies \cite{MG}.

\section*{Acknowledgements}

We have benefited from conversations with
Chryssomalis Chryssomalakos, Markus Deserno, Yair Guti\'errez Fosado, Eytan Katzav, Martin M\"uller and Eugene Starostin. Partial support from DGAPA PAPIIT grant IN114510 is acknowledged.

\begin{appendix}

\setcounter{equation}{0}
\renewcommand{\thesection}{Appendix \Alph{section}}
\renewcommand{\thesubsection}{A. \arabic{subsection}}
\renewcommand{\theequation}{A.\arabic{equation}}

\section{Tension in an unconstrained elastic space curve} \label{Stresstender}

There are various derivations of Eq. (\ref{CEEFSphere}). The approach
adopted here involves treating the tangent vector to the curve ${\bf
T}= {\bf Y}'$ as an   independent variable.  If the curve is also
parametrized by arc-length,  ${\bf T}$ is a unit vector, ${\bf
T}^2=1$. Thus, consider the energy functional ($\kappa^2 = {\bf
T}'^2$),
\begin{equation}
H_1 [{\bf Y}, {\bf T}, \Lambda, {\bf F}]
=  \int ds \, \left(\frac{1}{2}\, {\bf T}'{}^2 - \frac{1}{2}\Lambda ( {\bf T}^2
-1) + {\bf F}\cdot
({\bf T}- {\bf Y}')\right)\,;
\end{equation}
the presence of the two constraints liberates ${\bf T}$ to be varied
independently. Now the variation of $H_1$ with respect to${\bf Y}$
is given by
\begin{equation}
\delta_{\bf Y} H_1=  \int ds  \, {\bf F}'\cdot \delta  {\bf Y}\,,
\end{equation}
whereas its counterpart with respect to ${\bf T}$ is
\begin{equation}
\delta_{\bf T} H_1=  \int ds\, (  {\bf F} -   {\bf T}'' -   \Lambda\, {\bf T}
)\cdot \delta {\bf
T}\,.
\label{eq:Ft}
\end{equation}
Modulo the Frenet-Serret equations for the curve,  $\delta_{\bf T} H_1=0$ in
Eq. (\ref{eq:Ft}) implies
\begin{equation}
{\bf F}= (-\kappa^2 + \Lambda) \, {\bf T} + \kappa'\, {\bf N} + \kappa\tau\,{\bf
B}\,.
\end{equation}
For an isolated elastic curve, ${\bf F}'=0$. If the curve is constrained to lie
on a surface,  it was seen in Sec. \ref{confinedelastica} that ${\bf F}'=
\lambda {\bf n}$. Thus the tangential equation ${\bf t}\cdot {\bf F}' =0$
continues to hold. This equation can be integrated to determine the Lagrange
multiplier $\Lambda$ associated with the parametrization by arclength
\begin{equation}
\Lambda = \frac{3}{2}\kappa^2 - c\,,
\end{equation}
where $c$ is a constant of integration. Equation (\ref{EEFSforce}) follows.

\vskip1pc \noindent With respect to the Darboux frame, $\mathbf{F}$ is given by
\begin{equation} \label{eq:CEEF}
\mathbf{F} = \left(\frac{\kappa_g^2+\kappa_n^2}{2}  - c \right) \mathbf{T} +
\left(\kappa'_n +
\kappa_g \tau_g\right) \mathbf{n} + \left(\kappa_g'
-\kappa_n \tau_g\right) \mathbf{l}\,,
\end{equation}
where we have used the relations (\ref{dkappagn}) derived below.

\setcounter{equation}{0}
\renewcommand{\thesection}{Appendix \Alph{section}}
\renewcommand{\thesubsection}{B. \arabic{subsection}}
\renewcommand{\theequation}{B.\arabic{equation}}

\section{Darboux frame for surface curves} \label{AppDarbouxframe}

The structure equations (analogous  to the Frenet-Serret equations) describing
the Darboux frame are
given by
\begin{subequations}
\begin{eqnarray} \label{Darbouxequations}
\mathbf{T}' &=& \kappa_n \mathbf{n} + \kappa_g \mathbf{l}\,,\\
\mathbf{n}' &=& -\kappa_n \mathbf{T} - \tau_g {\bf l}\,, \\
\mathbf{l}' &=& -\kappa_g {\bf T} + \tau_g \mathbf {n}\,.
\end{eqnarray}
\end{subequations}
The normal curvature, the geodesic curvature, and torsion
are given, respectively, by
\begin{equation}
\kappa_n = {\bf T}'\cdot {\bf n} = -K_{ab} t^a t^b\,, \qquad
\kappa_g= {\bf T}' \cdot {\bf l}=  l^a t^b \nabla_b t_a
\,,\qquad
\tau_g = {\bf l}'\cdot{\bf n} = -K_{ab} t^a l^b \,.
\end{equation}
Here $t^a$ and $l^a$ are the components of the vectors ${\bf T}$ and ${\bf l}$
with respect to a
basis of surface tangent vectors adapted to the parametrization,  ${\bf e}_a$,
$a=1,2$: ${\bf T}=
t^a {\bf e}_a$, ${\bf l}= l^a {\bf e}_a$.
 $K_{ab}= {\bf e}_a\cdot \partial_b {\bf n}$ is the extrinsic curvature tensor
defined on $M$ and
$\nabla_a$ is the covariant derivative compatible with the induced surface
metric $g_{ab}= {\bf e}_a\cdot {\bf e}_b$. Whereas $\kappa_n$ and $\tau_g$
depend on the extrinsic curvature,  $\kappa_g$ is defined intrinsically; it
depends only on the surface metric $g_{ab}$. The Frenet curvatures are related
to their Darboux counterparts by
\begin{equation}\label{kaptau}
 \kappa_g = \kappa \, \sin \omega \,, \quad \kappa_n = \kappa \, \cos \omega \,,
\qquad \tau =
\omega' - \tau_g\,,
\end{equation}
so that $\kappa^2= \kappa_g^2 +\kappa_n^2$. Thus, the Frenet
curvature can be decomposed into its intrinsic and extrinsic parts.
The Frenet torsion $\tau$ is the sum of the derivative of the angle
connecting both frames and the geodesic torsion. Note that $\tau_g$
involves two derivatives, whereas $\tau$ involves three. The extra
derivative is associated with $\omega'$, the rotation rate of one
frame with respect to the other.

\vskip1pc \noindent
The identities Eqs. (\ref{kaptau})  imply the following:
\begin{equation} \label{dkappagn}
\kappa'_g =  \frac{\kappa'}{\kappa} \kappa_g + \kappa_n \left(\tau
+\tau_g\right)\,, \qquad
\kappa'_n =  \frac{\kappa'}{\kappa} \kappa_n - \kappa_g \left(\tau
+\tau_g\right)\,.
\end{equation}
These relations are used  in Sec. \ref{confinedelastica} to express the
Euler-Lagrange
derivatives for the curve in terms of $\kappa_g$, $\kappa_n$, and $\tau_g$ and
their derivatives.

\setcounter{equation}{0}
\renewcommand{\thesection}{Appendix \Alph{section}}
\renewcommand{\thesubsection}{C. \arabic{subsection}}
\renewcommand{\theequation}{C.\arabic{equation}}

\section{Identities for weak spherical confinement} \label{bcdetail}

We first derive the relationship between arc-length and the angle $\varphi$
correct to second order. One has
\begin{equation}
s = \int d\varphi \, \left[\left(\frac{d \vartheta} {d \varphi}\right)
^2 + \sin^ 2\vartheta \right]^{1/2}\,.
\end{equation}
Using Eq. (\ref {Eq:theta}) and the harmonic approximation for
$\kappa_g$ given by Eq. (\ref{linear}),  one obtains
\begin{equation}
\sin^2{\vartheta} \approx 1 - \frac{A_1^2}{M_0^2} \cos^2 n \varphi\,, \qquad
\mbox{and} \qquad
\frac{d \vartheta} {d \varphi} \approx \frac{1}{M_0} \, \frac{d \kappa_1 } {d
\varphi} = \frac{n A_1}{M_0} \,
\sin n \varphi \,.
\end{equation}
Thus, in this approximation,
\begin{equation} \label{skappa}
s \approx \varphi + \frac{1}{4} \frac{A_1^2}{(n^2-1)^2} \left((n^2-1) \varphi -
\frac{\left(n^2+1\right)}{2 n}\sin
   2 n \varphi \right)\,.
\end{equation}
This implies that\footnote{ $A_1$ vanishes when $n=1$, a solution which is
identified as a trivial rotation of the equatorial loop
about an axis in the equatorial plane.}
\begin{equation}
\label{eq:A1delR}
\Delta R   =\frac{1}{ 4 (n^2 -1)} \, A_1^2 \,.
\end{equation}
We now derive Eqs. (\ref{eq:delR1}) and (\ref{eq:delR2}). To do this
one needs to examine the boundary conditions (\ref{eq:c1}) and
(\ref{eq:c2}) correct to second order in $\epsilon=\sqrt{\Delta R}$.
At this order, the turning point $k_1$ of the potential
$V(\kappa_g)$, defined by Eq. (\ref{eq:Vdef}), coincides with $A_1$
given in the harmonic approximation by Eq.(\ref{eq:k1sM}). Thus, the roots in
the factorization of $E^2- V(\kappa)$ given in Eq. (\ref{eq:turningpoints}),
where $E$ is is the constant defined below Eq. (\ref{eq:Vdef}) are
\begin{equation}
k_1^2 = 2 (E-\sigma) \approx A^2_1\,, \qquad
K^2= 2 (E+\sigma) \approx - 4 n^2 - \frac {2}{n^2} (\sigma_2 - M_2) - 2 (
\sigma_2 +
M_2)\,.
\end{equation}
One thus has, correct to second order,
\begin{equation}
\frac{1}{\sqrt{E^2 -  V(\kappa_g)}} \approx
\frac{1}{n} \frac{1}{\sqrt{A_1^2 - \kappa_g^2}} \left(1 -
\frac {1}{4n^4} (\sigma_2 - M_2) -  \frac{1}{4n^2} ( \sigma_2 + M_2) -
\frac{1}{8 n^2}  \kappa_g^2 \right)\,.
\end{equation}
Equation (\ref{eq:c1}) then implies that
\begin{equation} \label{eq:delR1app}
\Delta R = - \frac {1}{4n^4} (\sigma_2 - M_2) -  \frac{1}{4n^2} ( \sigma_2 +
M_2) - \frac{1}{16 n^2}
A_1^2 \,.
\end{equation}
Furthermore, to second order,
\begin{equation}
M\,
\frac{ \frac{1}{2} \kappa_g^2 +
\sigma - 1}{ M^2 - \kappa_g^2} \approx
1 + \frac{1}{n^2-1}( \sigma_2 - M_2) + \frac{1}{2 (n^2-1)}\frac{n^2+1}{n^2-1}\,
\kappa_g^2\,.
\end{equation}
Equation (\ref{eq:c2}) then implies that for $n\ne 1$,
\begin{equation} \label{eq:delR2app}
\Delta R= -\frac{1}{n^2-1}( \sigma_2 - M_2) -
\frac{1}{4(n^2-1)}\frac{n^2+1}{n^2-1}\, A_1^2\,.
\end{equation}

\end{appendix}

\end{document}